Integrating sequencing datasets to form highly confident SNP and indel genotype calls for a whole human genome

Justin M. Zook[1,*], Brad Chapman[2], Jason Wang[3], David Mittelman[3,4], Oliver Hofmann[2], Winston Hide[2], Marc Salit[1]

[1]Biosystems and Biomaterials Division, National Institute of Standards and Technology, Gaithersburg, MD 20899
[2]Bioinformatics Core, Department of Biostatistics, Harvard School of Public Health, Cambridge, MA 02115
[3]Arpeggi, Inc, Austin, TX 78749
[4]Virginia Bioinformatics Institute and Department of Biological Sciences, Blacksburg, VA 24061
*jzook@nist.gov, 100 Bureau Dr, MS8313, Gaithersburg, MD 20899

Abstract:

Clinical adoption of human genome sequencing requires methods with known accuracy of genotype calls at millions or billions of positions across a genome. Previous work showing discordance amongst sequencing methods and algorithms has made clear the need for a highly accurate set of genotypes across a whole genome that could be used as a benchmark. We present methods to make highly confident SNP, indel, and homozygous reference genotype calls for NA12878, the pilot genome for the Genome in a Bottle Consortium. We minimize bias towards any method by integrating and arbitrating between 14 datasets from 5 sequencing technologies, 7 mappers, and 3 variant callers. Regions for which no confident genotype call could be made are identified as uncertain, and classified into different reasons for uncertainty. Our highly confident genotype calls are publicly available on the Genome Comparison and Analytic Testing (GCAT) website to enable real-time benchmarking of any method.

**Introduction:**

As whole human genome and targeted sequencing increasingly offer the potential to inform clinical decisions,[1-4] it is becoming critical to assess accuracy of variant calls and understand biases and sources of error in sequencing and bioinformatics methods.  Recent publications have demonstrated hundreds of thousands of differences between variant calls from different whole human genome sequencing methods or different bioinformatics methods.[5-11] These comparisons highlight the need for understanding variant call accuracy. In this report, we describe  a highly confident set of genome-wide genotype calls that can be used as a benchmark.  To minimize biases towards any sequencing platform or dataset, we compare and integrate 11 whole human genome and 3 exome datasets from five sequencing platforms for HapMap/1000 Genomes CEU female NA12878, which is a prospective Reference Material (RM) from the National Institute of Standards and Technology (NIST).  The recent approval of the first "Next Generation Sequencing" instrument by the FDA highlighted the utility of this candidate NIST RM in approving the assay for clinical use.[12]

NIST, with the Genome in a Bottle Consortium (www.genomeinabottle.org), is developing well-characterized whole genome RMs, which will be available for research, commercial, and clinical laboratories to sequence and assess variant call accuracy and understand biases. To create whole genome RMs, we need a best estimate of what is in each tube of RM DNA, describing potential biases and estimating the confidence of the reported characteristics.  To develop these data, we are developing methods to arbitrate between results from multiple sequencing and bioinformatics methods. The resulting arbitrated integrated genotypes can then be used as a benchmark to assess rates of false positive (FPs, or calling a variant at a homozygous reference site), false negatives (FNs, or calling homozygous reference at a variant site), and other genotype calling errors (e.g., calling homozygous variant at a heterozygous site).

FP rates are typically estimated by confirming a subset of variant calls with an orthogonal technology, which can be effective except for genome contexts that are also difficult for the orthogonal technology.[13] Genome-wide FN rates are much more difficult to estimate because the number of true negatives in the genome is overwhelmingly large (*i.e.*, most bases match the reference assembly).  Typically, if studies estimate FN rates, they use microarray data from the same sample, but microarrays are hypothesis-driven, in that they only have genotype content with known common SNPs in regions of the genome accessible to the technology.  Transition/transversion (Ti/Tv) ratios are sometimes used to estimate FP rates for SNPs, since lower values closer to 0.5 tend to indicate more FPs.  However, Ti/Tv of real mutations is variable for different regions of the genome (e.g., exome vs. non-exome), and the "expected" Ti/Tv is not clear since it is also hypothesis-driven, derived empirically from "easier" and more common variants.

Therefore, we propose the use of well-characterized whole genome reference materials to estimate both FN and FP rates of any sequencing method, as opposed to using one orthogonal method that may have correlated biases in genotyping and "blind spots," or an emphasis on common variants and more accessible regions of the genome.  When characterizing the reference material itself, both a low FN rate (*i.e.*, calling a high proportion of true variant genotypes, or high sensitivity) and a low FP rate (*i.e.*, a high proportion of the called variant genotypes are correct, or high specificity) are very important (see Supplementary Table S1).  A low FN rate is important because locations that are incorrectly called non-variant in the highly confident call set would cause the FP rate of an accurate method to be overestimated. Conversely, a low FP rate is important because locations that are incorrectly called variant in the highly confident call set would cause the FN rate of an accurate method to be overestimated. Bases with uncertain genotypes in the highly confident call set are not as detrimental to performance assessment as incorrect genotypes. However, uncertain genotypes in the reference material will typically result in lower estimated FN and FP rates for a method being examined, because they tend to fall in difficult-to-sequence regions of the genome.

Low FP and FN rates cannot be reliably obtained from filtering of low variant quality scores alone, because biases in the sequencing and bioinformatics methods are not all included in the variant quality scores. Therefore, several variant callers use a variety of characteristics (or annotations) of variants to filter likely FP calls from a dataset. However, filtering FPs without filtering true variants can be difficult, since annotations are not perfectly specific for errors.

While large datasets such as whole genome sequencing datasets pose challenges for analysis due to their large size, machine learning algorithms can take advantage of the large number of sites across a whole human genome to learn the optimal way to combine annotations and filter genotype errors. For example, the Genome Analysis ToolKit (GATK) includes a Variant Quality Score Recalibration (VQSR) module that uses annotations related to strand bias, mapping quality, allele balance, position inside the read, etc. to filter potential errors.[14, 15] GATK trains a Gaussian Mixture Model using suspected true positive variants to find the optimal way to filter FPs while retaining a specified sensitivity to likely true positive variants. The current GATK Best Practices (v4) recommend using HapMap sites as likely true positives, and OMNI microarray and HapMap sites for training. We have adapted GATK VQSR to use sites that are mostly concordant across multiple sequencing datasets from different platforms as the training set to give it a larger, more robust training set.

Currently, GATK and other variant callers do not effectively use multiple datasets from the same sample to refine genotype calls and find likely FPs and FNs. A couple methods have recently been proposed by the 1000 Genomes Project to integrate multiple variant call sets, but these methods have not been used to arbitrate between datasets from different sequencing methods on the same genome.[16] Therefore, we have extended GATK's methods to integrate information from multiple publicly available datasets of the same sample, and use VQSR to identify possible FPs and arbitrate between discordant datasets (see Fig. 1 and Supplementary Fig. S1). In this way, we can integrate datasets from multiple sequencing technologies and minimize bias towards any particular sequencing technology, similar to co-training in semi-supervised machine learning.[17] Any particular technology has locations in the genome that are erroneous or ambiguous due to systematic sequencing errors (e.g., the GGT motif in Illumina) or biases in coverage.[7] While algorithms can filter some systematic sequencing errors, they cannot perfectly distinguish between false positives and true positives. Using multiple platforms that have different systematic sequencing errors can help distinguish between true positives and false positives in any particular sequencing technology. In addition, regions with very low coverage in one platform can have sufficient coverage in a different platform to make a genotype call. The resulting methods, RMs, and integrated genotype calls are a public resource at www.genomeinabottle.org for anyone to characterize performance of any genome sequencing method.

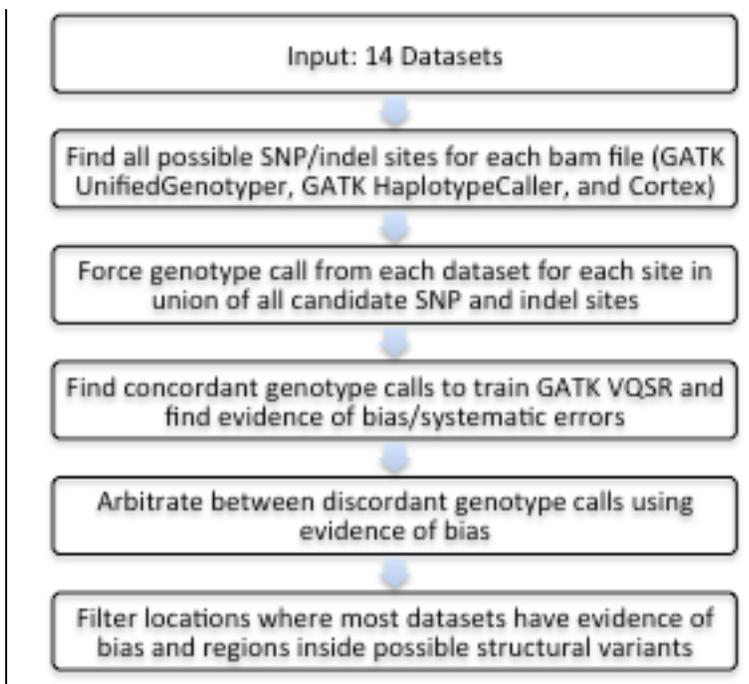

Fig. 1: Description of the process used to develop highly confident genotype calls by arbitrating differences between multiple datasets from different sequencing platforms, and define regions of the genome that could be confidently genotyped.  A more detailed description of our methods and examples of arbitration are in Supplementary Figs. S1-S3.

**Results**
*Arbitrating between datasets that disagree*
To develop our highly confident genotype calls, we use the 11 whole genome and 3 exome datasets from 5 sequencing platforms and 7 mappers, as described in Table 1.  For the hundreds of thousands of possible SNP sites in the whole genome that differ between sequencing platforms and variant callers, we developed methods to identify biases and arbitrate between differing datasets (see Fig. 1 and Supplementary Figs. S1 to S3).   Briefly, we first selected all sites that had even small evidence for a SNP or indel in any dataset.  Then, we used previously existing and new annotations in the GATK VQSR Gaussian Mixture Model to identify sites in each dataset with characteristics of biases, including systematic sequencing errors (SSEs),[18, 19] local alignment difficulties, mapping difficulties, or abnormal allele balance.  Unlike the normal VQSR methods, we trained VQSR independently for each dataset for homozygous and heterozygous calls using consensus genotypes across all datasets.  For each site where genotypes in different datasets disagreed, we sequentially filtered datasets with characteristics of (1) systematic sequencing errors, (2) alignment uncertainty, and (3) atypical allele balance.  We progressively filtered lower VQSR tranches of each characteristic until at least 5 times more datasets agree than disagree (e.g., if 5 or more datasets confidently call one genotype, and 1 or fewer datasets confidently call a different genotype).  If fewer than 2 remaining datasets agree, or if the remaining datasets had characteristics of systematic sequencing errors, local alignment difficulties, mapping difficulties, or abnormal allele balance, then the site was considered uncertain.  Note that mapping bias was only used to mark sites as uncertain because mapping quality scores are not scaled similarly to allow arbitration between datasets mapped with different algorithms.  In addition, we filter as uncertain (1) regions with simple repeats not completely covered by reads from any dataset, (2) regions with known tandem duplications not in the GRCh37 genome reference assembly (which was partly developed from NA12878 fosmid clones and is available at http://humanparalogy.gs.washington.edu/), (2) regions paralogous to the 1000 Genomes "decoy reference", (3) regions in the RepeatSeq database, and (4) regions inside structural variants for NA12878 that have been submitted to dbVar.  We provide a file in bed format that specifies the regions in which we can confidently call the genotype.  As shown in Table S2, before filtering structural variants, we are able to call confidently 87.6% of the non-N bases in chromosomes 1-22 and X, including 2,484,884,293 homozygous reference sites, 3,137,725 SNPs, and 201,629 indels. Excluding structural variants conservatively excludes an additional 10% of the genome, with 2,195,078,292 homozygous reference sites, 2,741,014 SNPs, and 174,718 indels remaining.  The bed file containing structural variants, as well as some of the other bed files containing uncertain regions, can also be used to help identify sites in an assessed variant call file that may be questionable.  All vcf and bed files are publicly available on the Genome in a Bottle ftp site described in the Online Methods.

    We also varied the cut-offs used to differentiate between highly confident and uncertain variants, and found that they had only a small effect.  When varying the number of supporting datasets required to make a highly confident genotype call between 1 and 3, less than 0.05 % of the variants were lost or gained compared to requiring 2 datasets.  This very small change likely results partially from other requirements, such as our condition that highly confident sites have at least 2 datasets without evidence of bias.  Another parameter we varied was the ratio of datasets that agree with the genotype to datasets that have a different genotype.  With a ratio of 8 or 3, the calls differ by less than 0.5% compared to a ratio of 5.  Most of the calls lost or gained were concordant with the fosmid calls, and most of the discordant sites were correct in our highly confident genotypes.  However, a small number of sites gained by lowering thresholds appeared to be questionable or possible false positives.  In general, varying cutoffs has very little effect on the resulting genotypes, demonstrating the robustness of our methods that integrate multiple datasets.

Table 1: Description of datasets and their processing to produce bam files for our integration methods

| Source* | Platform | Mapping algorithm | Coverage | Read length | Genome/exome |
|---|---|---|---|---|---|
| 1000 Genomes | Illumina GaIIx | Bwa | 39 | 44 | Genome |
| 1000 Genomes | Illumina GaIIx | Bwa | 30 | 54 | Exome |
| 1000 Genomes | 454 | Ssaha2 | 16 | 239 | Genome |
| X Prize | Illumina HiSeq | Novoalign | 37 | 100 | Genome |
| X Prize | SOLiD 4 | Lifescope | 24 | 40 | Genome |
| Complete Genomics | Complete Genomics | CGTools 2.0 | 73 | 33 | Genome |
| Broad | Illumina HiSeq | Bwa | 68 | 93 | Genome |
| Broad | Illumina HiSeq | Bwa | 66 | 66 | Exome |
| Illumina | Illumina HiSeq | CASAVA | 80 | 100 | Genome |
| Illumina | Illumina HiSeq – PCR-free | Bwa | 56 | 99 | Genome |
| Illumina | Illumina HiSeq – PCR-free | Bwa | 190 | 99 | Genome |
| Life Technologies | Ion Torrent | tmap | 80 | 237 | Exome |
| Illumina | Illumina HiSeq – PCR-free | BWA-MEM | 60 | 250 | Genome |
| Life Technologies | Ion Torrent | tmap | 12 | 200 | Genome |

*These data and other datasets for NA12878 are available at the Genome in a Bottle ftp site at NCBI (ftp://ftp-trace.ncbi.nih.gov/giab/ftp/data/NA12878) and are described on a spreadsheet at http://genomeinabottle.org/blog-entry/existing-and-future-na12878-datasets.

*Different variant representations make comparison difficult*

Indels and complex variants (i.e., nearby SNPs and indels) are particularly difficult to compare across different variant callers, because they can frequently be represented correctly in multiple ways. Therefore, we first regularized each of the variant call files (vcf) using the vcfallelicprimitives module in vcflib (https://github.com/ekg/vcflib). Regularization minimizes counting different methods of expressing the same variant (e.g., nearby SNPs/indels) as different variants. Our regularization procedure splits adjacent SNPs into individual SNPs, left-aligns indels, and regularizes representation of homozygous complex variants. However, it cannot regularize heterozygous complex variants without phasing information in the vcf, such as individuals that are heterozygous for the CAGTGA>TCTCT change that is aligned in 4 different ways in Fig. 2. Most alignment-based variant callers would output 4 different vcf files for these 4 representations. To avoid this problem in our integration process, our calls are represented in the output format of GATK HaplotypeCaller 2.6-4, which regularizes representation by performing de novo assembly. When comparing calls from other variant callers, we recommend using the vcfallelicprimitives module in vcflib, as well as manual inspection of some discordant sites to determine whether the calls are actually different representations of the same complex variant.

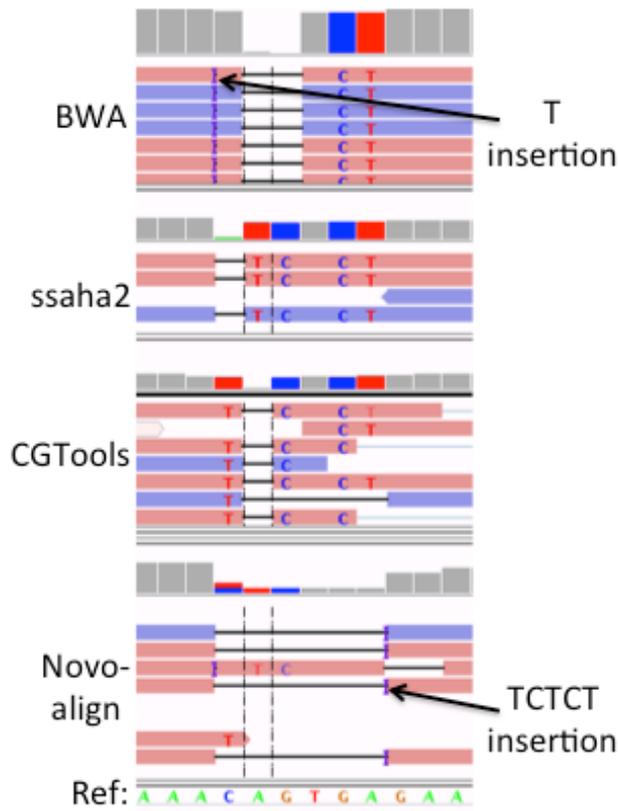

Fig. 2: Example of complex variant with 4 different representations from 4 different mappers, which can cause datasets to appear to call different variants when in reality they are the same variant. In this case, the 6 bases CAGTGA are replaced by the 5 bases TCTCT at location 114841792-114841797 on chromosome 1. The 4 sets of reads are from Illumina mapped with BWA, 454 mapped with ssaha2, Complete Genomics mapped with CGTools, and Illumina mapped with Novoalign.

*Integrated variant calls are highly sensitive and specific*

Transition/transversion ratio (Ti/Tv) is sometimes used as a metric for accuracy of calls, since the biological Ti/Tv is significantly higher than the 0.5 Ti/Tv expected from random sequencing errors. As shown in Table 2, our integrated calls have a Ti/Tv comparable to the other datasets for common variants in the whole genome and exome, but our integrated calls have a higher Ti/Tv than the other datasets for novel whole genome variants, which usually indicates a lower error rate. However, it should be noted that Ti/Tv is limited in its use since the assumption that novel or more difficult variants should have the same Ti/Tv as common variants may not be true.[20]

Table 2: Performance assessment of 250bp Illumina sequencing mapped with BWA-MEM and called with GATK HaplotypeCaller v2.6 (250bp_HC), Complete Genomics sequencing from 2010 (CG), and our integrated calls vs. OMNI microarray SNPs and vs. our Integrated SNPs/indels, as well as their overall transition/transversion ratio (Ti/Tv)

| Dataset | Capture? | OMNI SNPs with Integrated BED file | | OMNI SNPs without BED file | | Integrated SNPs with BED file | | Integrated indels with BED file | | Common Variants | Novel Variants |
|---|---|---|---|---|---|---|---|---|---|---|---|
| | | Sensitivity | Specificity | Sensitivity | Specificity | Sensitivity | PR* | Sensitivity | PR* | Ti/Tv | Ti/Tv |
| 250bp_HC | Genome | 99.49% | 99.97% | 98.47% | 99.93% | 99.90% | 99.73% | 99.55% | 93.11% | 2.04 | 1.43 |
| CG | Genome | 98.55% | 99.98% | 97.11% | 99.96% | 97.09% | 99.27% | 72.27% | 89.43% | 2.10 | 1.29 |
| Integrated | Genome | 99.54% | 99.98% | n/a | n/a | n/a | n/a | n/a | n/a | 2.14 | 1.94 |
| 250bp_HC | Exome | 99.55% | 99.98% | 99.10% | 99.96% | 99.90% | 99.58% | 100.00% | 94.60% | 2.60 | 1.57 |
| CG | Exome | 98.35% | 99.99% | 97.64% | 99.96% | 99.00% | 99.04% | 90.00% | 85.86% | 2.71 | 1.04 |
| Integrated | Exome | 99.57% | 99.98% | n/a | n/a | n/a | n/a | n/a | n/a | 2.92 | 1.33** |

* Precision ratio (PR) = TP/(TP+FP) – note that Specificity of all datasets vs. our integrated calls is 100.00% due to the large number of TNs.
** Our integrated calls only contain 30 novel variants in the exome, so the Ti/Tv has a high uncertainty

Genotyping microarrays are an orthogonal measurement method that is sometimes used to assess the accuracy of sequencing genotype calls at sites interrogated by the microarray.[13] When assessed against microarray genotype calls, our integrated genotype calls are highly sensitive and specific (see Table 2). We correctly called 564,410 SNPs on the microarray, and there were 1,332 SNPs called by the microarray not in our calls, and 527 variants calls in our set that were at positions called homozygous reference in the microarray. We manually inspected 2 % of the SNPs specific to the microarray and 4 % of the calls specific to our calls. For the manually inspected SNPs specific to the microarray, about half were clearly homozygous reference in all sequencing datasets, without any nearby confounding variants (see Supplementary Fig. S19). In addition, several sites were adjacent to homopolymers, which were mostly correctly called indels in our highly confident calls (see Supplementary Figs. S20 and S21), but a couple of our calls had incorrect indel lengths (see Supplementary Figs. S22 and S23). A few sites were also incorrectly called SNPs by the array that may have been confounded by nearby variants (Supplementary Figs. S24 and S25). We also manually inspected our calls at locations that were homozygous reference in the microarray, and found that many were true indels. These calls were either true indels (see Supplementary Fig. S26), had nearby variants that confounded the probe (see Supplementary Fig. S27), or the probe was interrogating a different SNP than this sample had (see Supplementary Fig. S28).

We also compared our SNP and indel calls to "high quality variants" found in multiple sequencing platforms (mostly sequenced using Sanger sequencing) for the GeT-RM project (http://www.ncbi.nlm.nih.gov/variation/tools/get-rm/). Our integrated calls correctly genotyped all 427 SNPs and 42 indels. In addition, we compared to Sanger sequencing data from the XPrize, and found that the calls were concordant for all 124 SNPs and 37 indels. In addition, we determined that none of our highly confident variants fall in the 366,618 bases that were covered by high quality homozygous reference Sanger reads from the GeT-RM project.

In addition, to understand the accuracy of both our SNP and indel calls across larger regions of the genome, we compared our calls to the fosmid calls generated by the XPrize from Illumina and SOLiD sequencing of fosmids covering one allele of ~5% of the genome. Fosmid sequencing is advantageous in that only one allele is measured, so no heterozygous genotypes should exist. However, because only one allele is measured, it can assess both FP and FN rates of homozygous calls, but it can only assess FN rates of heterozygous calls in our integrated calls. Our calls were highly concordant overall, with 76,698 concordant homozygous SNP calls, 58,954 concordant heterozygous SNP calls, 5,061 concordant homozygous indel calls, and 5,881 concordant heterozygous indel calls.

To understand which method was correct when our integrated calls and the fosmid calls were discordant, we manually curated alignments from several datasets in the regions around a randomly selected 25% of the discordant variants (see the more detailed discussion in the Supplementary information and Supplementary Figs. S4 to S17 for some examples). Manual curation of alignments from multiple datasets, aligners, and sequencing platforms allowed us to resolve the reasons for all of the differences between our integrated calls and the fosmid calls. For the manually curated variants in our integrated calls and not in the fosmid calls, almost all were FNs in the fosmid calls due to mis-called complex variants (see Supplementary Fig. S4) or overly stringent filtering (see Supplementary Fig. S5). For variants in the fosmid calls and not in our integrated calls, several reasons were found for the differences, but our integrated calls appeared to be correct except for a few partial complex variant calls. The fosmids contain 119 million reference bases in our highly confident regions, and we manually curated 25% of discordant variants in these bases. Therefore, this analysis suggests that our integrated calls likely contain ~3 partial complex variant calls and between 0 and 1 false positive or false negative simple SNP or indel calls per 30 million highly confident bases, in which our integrated calls contain ~94,500 TP SNPs and ~1400 TP indels.

Finally, to ensure our variant calling methods are not missing any sites that might be found by other variant callers, we compared our highly confident genotypes to an callset generated by freebayes on Illumina exome sequencing data. There were 208 variants in the freebayes variant calls with coverage greater than 20 that we called highly confident homozygous reference. We manually inspected a random 10 % of these putative variants, and all of them appeared to be likely false positives due to systematic sequencing or mapping errors, or sites where freebayes called an inaccurate genotype for the correct variant.

*Assessing performance of SNP sequencing vs. arrays overestimates sensitivity for whole genome calls*

While microarrays can be useful to help understand sequencing performance, they can only assess performance in the regions of the genome accessible to microarrays (i.e., sequences to which probes can bind and bind uniquely). In addition, microarray genotypes can be confounded by nearby phased variants that are not in the array probes. Microarrays tend to contain known common SNPs and avoid genome regions of low complexity. For example, if "low complexity" is defined as having genome mappability scores[21] less than 50 for Illumina, SOLiD, or Ion Torrent, then only 0.0117 % of the microarray sites are in low complexity regions, compared to 0.7847 % of integrated variants. Our integrated calls have a 67 times higher percentage of low complexity regions compared to microarrays, but even our integrated calls ignore many regions of low complexity since 9.8% of uncertain sites have low complexity. To understand the impact of including lower complexity sites for performance assessment, we explored the use of our integrated genotype calls as a benchmark to assess genotype calls from single datasets, and compared this assessment to an assessment using microarrays (see Table 2). Many of the sites that are discordant in the microarray are due to errors in the microarray, as we show in Supplementary Figs. S19 to S28, so FP rates are actually lower when sequencing datasets are assessed against our highly confident genotypes. Apparent FN rates are approximately the same, but many of the apparent FNs when assessed against the microarray are actually FPs in the microarray.

*Using highly confident calls to understand and improve performance*
As an example, we selected a new whole genome variant call set from the Broad Institute/1000 Genomes Project to show how this set of highly confident genotype calls can be used to understand and improve performance even for new versions of a sequencing technology (2x250 paired-end Illumina reads), mapping algorithm (bwa-mem),[22] and variant caller (GATK v.2.6 HaplotypeCaller).[15] In addition, we compared an older Complete Genomics callset to see how calls from a completely different pipeline compare. We also assessed performance of several exome datasets on GCAT that use 150x Illumina+Novoalign+Freebayes,[23] Illumina+Novoalign+Samtools,[24] Illumina+bwa+Freebayes,[25] and 30x Ion Torrent+Tmap+GATK-HaplotypeCaller.

Fig. 3 and Supplementary Figs. S31, S32, and S35 contain ROC curves showing how FP and TP rate change while varying the cutoff for read depth or variant quality score (not applicable to Complete Genomics). Variant quality score gives a better ROC curve than read depth in most cases, likely because sites with very high read depth can actually have higher error rates due to mapping problems. The new 250-bp Illumina whole genome with HaplotypeCaller has a higher accuracy than the older Complete Genomics or any of the exome sequencing datasets for both SNPs and indels. The 150x Illumina exome callsets have a higher accuracy than the 30x Ion Torrent exome callset, particularly for indels. The accuracy for SNPs is much higher than the accuracy for indels in all callsets, which is expected since indels are more difficult to detect than SNPs, especially in homopolymers and low complexity sequence. From the ROC curves, it is apparent that the variant quality score cutoff for the HaplotypeCaller for this dataset is probably not optimal, since raising the cutoff could significantly lower the FP rate while only minimally increasing the FN rate.

Direct observation of alignments around discordant genotypes is often a useful way to understand the reasons behind inaccurate genotype calls. For example, Fig. S37 shows an example of an apparent systematic Illumina sequencing error that is in both the new HaplotypeCaller and UnifiedGenotyper callsets, but arbitrated correctly in the integrated callset. Many of the differences are due to difficult regions with low mapping quality, where it is often difficult to determine the correct answer from short read sequencing (e.g., Fig. S38).

The variant calls in any dataset can also be intersected with our bed files containing different classes of "difficult regions" of the genome, as shown in Table S3. These comparisons can identify potentially questionable variant calls that should be examined more closely. About 1 million variants called in the 250 bp Illumina HaplotypeCaller vcf are inside NA12878 structural variants reported to dbVar, which is the largest number of variants in any category.

Further work will need to be done to determine which structural variants are accurate, but variants in these regions could be inspected further. There are also over 200,000 variants called in the 250 bp Illumina HaplotypeCaller vcf in each of several uncertain categories: sites with unresolved conflicting genotypes, known segmental duplications, regions with low coverage or mapping quality, and simple repeats. While many of these variants may be true variants, they could be examined more closely to identify potential FPs.

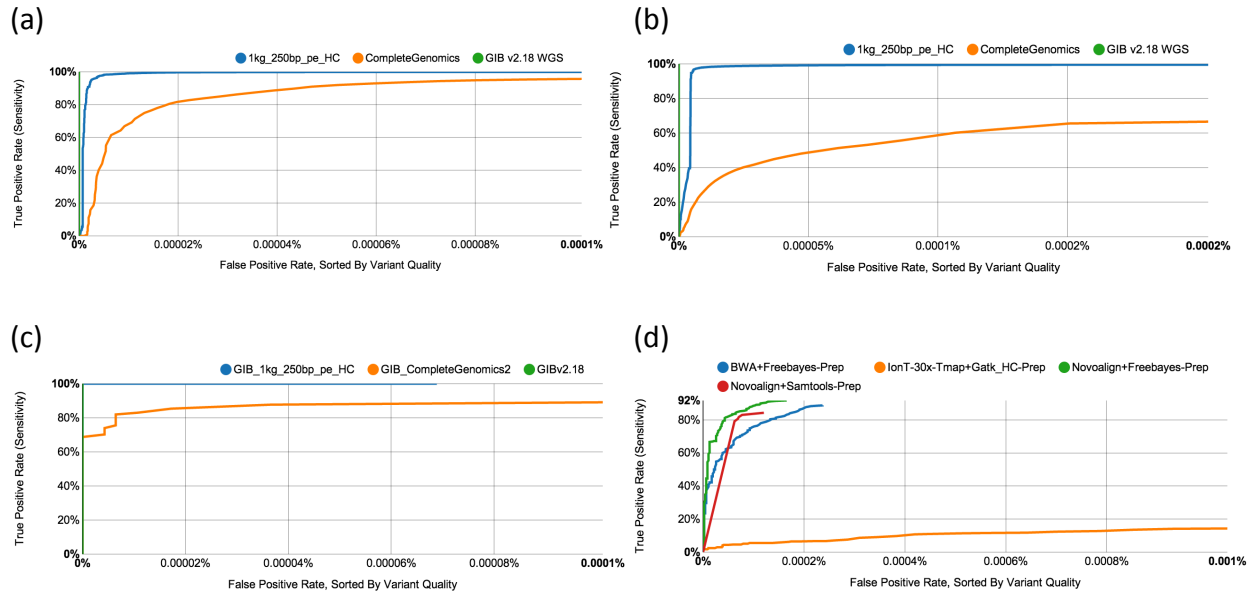

Fig. 3: Receiver Operating Characteristic (ROC) curves plotting True Positive Rate (sensitivity) vs. False Positive Rate, with variants sorted by variant quality score, for (a) whole genome SNPs, (b) whole genome indels, and (c-d) exome indels. The assessed variant calls come from Complete Genomics 2.0 (CompleteGenomics2 and GiB_CompleteGenomics), 250bp Illumina mapped with BWA-MEM and called with GATK HaplotypeCaller v2.6 (1kg_250bp_pe_HC and GiB_1kg_250bp_pe_HC), 150x Illumina exome sequencing mapped with BWA and called with Freebayes (BWA+Freebayes-Prep), 30x Ion Torrent exome sequencing mapped with Tmap and called with GATK HaplotypeCaller (IonT-30x-Tmap+Gatk_HC-Prep), 150x Illumina exome sequencing mapped with Novoalign and called with Freebayes (Novoalign+Freebayes-Prep), and 150x Illumina exome sequencing mapped with Novoalign and called with Samtools (Novoalign+Samtools-Prep).

**Discussion**

To develop a benchmark whole genome dataset, we have developed the first set of methods to integrate sequencing datasets from multiple sequencing technologies to form highly confident SNP and indel genotype calls. The resulting genotype calls are more sensitive and specific and less biased than any individual dataset, because our methods use characteristics of biases associated with systematic sequencing errors, local alignment errors, and mapping errors in individual datasets. We also minimize bias towards any individual sequencing platform by requiring that at least 5 times more datasets agree than disagree, so that all 10 datasets would have to agree if 2 had a different genotype. Therefore, even though there are more Illumina datasets, other platforms would have to agree with the Illumina datasets for them to override 2 datasets that disagreed.  In addition, we include an annotation "platforms" in the INFO field that specifies the number of platforms that support a call. If a user would like to minimize any potential platform bias even further, they can select only variants that have support in 2 or more platforms.

It is critical to understand that the process used to generate any set of benchmark genotype calls can affect the results of performance assessment in multiple ways, as depicted in Supplementary Table S1: (1) If many "difficult" regions of genome are excluded from the truth dataset (or labeled "uncertain," meaning that they may be down-weighted or disregarded in performance assessment), any assessed datasets will have lower apparent FP and FN rates than if the difficult regions were included.  Therefore, it is important to recognize that any comparison to our highly confident genotypes excludes the most difficult variants and regions of the genome (currently ~23 % of the genome, including potential structural variants, see Supplementary Table S2). In addition, almost all of the indels are currently < 40 bp in length.  (2) Any FP variant calls in the truth dataset could result in an assessed FN rate higher than the true FN rate if the assessed calls are correct, or in an assessed FP rate lower than the true FP rate if the assessed calls are also FPs.  (3) Any FN variant calls in the truth dataset could result in an assessed FP rate higher than the true FP rate if the assessed calls are correct, or in an assessed FN rate lower than the true FN rate if the assessed calls are also FNs.  (4) Many comparison tools treat heterozygous and homozygous variant genotype calls as equivalent, which enables simple calculations of sensitivity and specificity, but these genotypes can have different phenotypes, so it is often important to assess whether the genotype is accurate, as we do in this work, not just whether a variant is detected.

In general, for the benchmark calls to be useful for performance assessment, the FP rate of the benchmark should be much lower than the FN rate of the assessed dataset, and the FN rate of the benchmark should be much lower than the FP rate of the assessed dataset. To be confident our benchmark integrated calls are not biased toward any sequencing or bioinformatics method and have sufficiently low FN and FP rates, we compared our integrated calls to multiple independent methods (microarrays, capillary sequencing, fosmid sequencing, Illumina exome sequencing called with freebayes, and new 2x250bp long-read Illumina sequencing mapped with a new algorithm BWA-MEM and analyzed with a new version of GATK).

While we have shown our integrated calls have very low FP and FN rates, we recommend that users of our integrated calls examine alignments around a subset of discordant genotype calls, such as using the new GeT-RM browser for NA12878 (http://www.ncbi.nlm.nih.gov/variation/tools/get-rm/). Overall statistics like sensitivity and specificity are useful, but manual inspection of alignments from multiple datasets (or even a single dataset) for a subset of discordant sites is essential for properly understanding any comparison between variant call sets. Manual inspection can help identify discordant representations of the same variant, potential biases in sequencing/bioinformatics methods, and difficult regions of the genome where variant calls may be questionable.  In addition to comparing genotype calls in the regions for which we determined we can make confident integrated genotype calls, we also recommend examining variant calls in regions we consider uncertain for different reasons.  Examining these difficult regions can help identify variants that may be questionable. We also encourage contacting the authors of this manuscript if any genotypes in our integrated calls are questionable or in error, as this call set will be maintained and refined over time as new sequencing and analysis methods become available.

These methods represent the basis of methods to form highly confident genotype calls for genomes selected as RMs by the new Genome in a Bottle Consortium. This consortium will develop the reference materials, reference data, and reference methods to help enable translation of genome sequencing to clinical practice.  As we show in this work, highly confident genotype calls from a well-characterized whole genome are useful for assessing biases and rates of accurate and inaccurate genotype calls in any combination of sequencing and bioinformatics methods.  Highly confident genotype calls for publicly available genomes will be particularly useful for performance assessment of rapidly evolving sequencing and bioinformatics methods.  This resource is publicly available through the Genome in a Bottle Consortium website ([www.genomeinabottle.org](www.genomeinabottle.org)) and ftp site at NCBI ([ftp://ftp-trace.ncbi.nih.gov/giab/ftp/data/NA12878)](ftp://ftp-trace.ncbi.nih.gov/giab/ftp/data/NA12878)).  The resource will continue to be improved and updated with additional types of variants (e.g., complex variants, and structural variants) and with increasingly difficult regions of the genome, incorporating new sequencing data as it is collected.

**Acknowledgments**
We thank Justin Johnson and Anjana Varadarajan from the Archon Genomics X Prize and EdgeBio for contributing their whole genome sequencing data from SOLiD and Illumina, Complete Genomics and Life Technologies for providing bam files for NA12878, and the Broad Institute and 1000 Genomes Project for making publicly available bam and vcf files for NA12878.  The Illumina exome data on GCAT was given to the Mittelman lab by Dr. Michael Linderman at Icahn Institute of Genomics and Multiscale Biology of the Icahn School of Medicine at Mount Sinai. Certain commercial equipment, instruments, or materials are identified in this document. Such identification does not imply recommendation or endorsement by the National Institute of Standards and Technology, nor does it imply that the products identified are necessarily the best available for the purpose.

**Online Methods**

*Datasets*

Nine whole genome and three exome sequencing datasets (see Table 1 for details about source, platform, mapping algorithm, coverage, and aligned read length) were used to form the integrated genotype calls for Coriell DNA sample NA12878. Six whole genome (two PCR-free) and two exome datasets were from Illumina sequencers, one whole genome from SOLiD sequencers, one whole genome from 454 sequencer, one whole genome from Complete Genomics, and one exome from Ion Torrent.[26] Some have bam files publicly available, which were used directly in this work. These data and other datasets for NA12878 are available at the Genome in a Bottle ftp site at NCBI (ftp://ftp-trace.ncbi.nih.gov/giab/ftp/data/NA12878) and are described on a spreadsheet at http://genomeinabottle.org/blog-entry/existing-and-future-na12878-datasets. In addition, the results of this work (highly confident variant calls and bed files describing confident regions) are available at ftp://ftp-trace.ncbi.nih.gov/giab/ftp/data/NA12878/variant_calls/NIST along with a README.NIST describing the files and how to use them. The files used in this manuscript are NISTIntegratedCalls_14datasets_131103_HetHomVarPASS_VQSRv2.18_2mindatasets_5minYesNoRatio_all_nouncert_excludesimplerep_excludesegdups_excludedecoy_excludeRepSeqSTRs_noCNVs_vardist.vcf.gz, which contains highly confident heterozygous and homozygous variant calls, and union13callableMQonlymerged_addcert_nouncert_excludesimplerep_excludesegdups_excludedecoy_excludeRepSeqSTRs_noCNVs_v2.18_2mindatasets_5minYesNoRatio.bed.gz, which contains intervals that can be considered highly confident homozygous reference (for snps and short indels) if there is not a variant in the vcf.

*Comparison of variant calls using different methods*

To compare variants called using different methods, we first sought to normalize representation of short indels, complex variants, and multinucleotide polymorphisms (MNPs) so that the same variant represented in different ways would not be considered discordant. We used the vcfallelicprimitives module in vcflib (https://github.com/ekg/vcflib) to help regularize representation of these variants. Regularization minimizes counting different methods of expressing the same variant (e.g., nearby SNPs/indels) as different variants. Our regularization procedure splits adjacent SNPs into individual SNPs, left-aligns indels, and regularizes representation of homozygous complex variants. However, it cannot regularize heterozygous complex variants without phasing information in the vcf, such as individuals that are heterozygous for the CAGTGA>TCTCT change that is aligned in 4 different ways in Fig. 2. Regularizing heterozygous complex variants without phasing information is not generally possible because they could be phased in multiple ways. All other shell (Sun Grid Engine) and perl scripts written for this work and the resulting bed file are publicly available at https://github.com/jzook/genome-data-integration.

*Obtaining highly confident genotypes for training VQSR*

To reduce the number of sites that need to be processed, we first used GATK (v. 2.6-4) UnifiedGenotyper and HaplotypeCaller with a low variant quality score threshold of 2 to find all possible SNP and indel sites in each dataset except Complete Genomics. For Complete Genomics, we used their unfiltered set of SNP calls from CGTools 2.0. In addition, we included sites called by Cortex *de novo* assembly method for the ~40x Illumina PCR-free dataset. The union of these sites from all datasets served as our set of possible SNP sites for downstream processing.

Since each dataset did not make a genotype call at every possible SNP and indel site, we forced GATK UnifiedGenotyper to call genotypes for each dataset individually at all of the possible SNP sites (GATK_..._UG_recall_...sh). In addition, we forced GATK HaplotypeCaller to perform local *de novo* assembly around all candidate indels and complex variants for each dataset individually (GATK_..._haplo_recall…sh). We then combined the UG and HC calls, giving preference to HC within 20bp of an HC indel with a PL>20. We used the genotype likelihoods (PL in vcf file) to determine which sites had genotypes confidently assigned across multiple datasets. We used the minimum non-zero PL (PLdiff), which is the Phred-scaled ratio of the likelihoods of the most likely genotype to the next most likely genotype (similar to the Most Probable Genotype described previously[27]). In addition, we divided PLdiff by the depth of coverage (PLdiff/DP) as a surrogate for allele balance because PLdiff should increase linearly with coverage in the absence of bias. For a heterozygous variant site to be used to train VQSR, we required that PLdiff>20 for at least 2 datasets, the net PLdiff for all datasets > 100, the net PLdiff/DP for all datasets > 3.4, fewer than 15% of the datasets had PLdiff>20 for a different genotype, fewer than 30% of the datasets have >20% of the reads with mapping quality zero, and fewer than 2 datasets have a nearby indel called by HaplotypeCaller but do not call this variant. For a homozygous variant site to be used to train VQSR, we required that PLdiff>20 for at least 2 datasets, the net PLdiff for all datasets > 80, the net PLdiff/DP for all datasets > 0.8, fewer than 25% of the datasets had PLdiff>20 for a different genotype, and fewer than 2 datasets have a nearby indel called by HaplotypeCaller but do not call this variant. These requirements were specified to select generally concordant sites with reasonable coverage and allele balances near 0, 0.5, or 1 for training VQSR.

These highly concordant heterozygous and homozygous variant genotypes were used independently to train the VQSR Gaussian Mixture Model separately for each dataset for heterozygous and homozygous (variant and reference) genotypes. Unlike the normal VQSR process, we train on heterozygous and homozygous genotypes independently because they could have different distributions of annotations and different characteristics of bias. We fit only a single Gaussian distribution to each annotation since most of the annotations have approximately Gaussian distributions. Thus, additional Gaussians often fit noise in the data, and the model frequently does not converge when attempting to fit more than one Gaussian. We fit VQSR Gaussian mixture models for annotations associated with alignment problems, mapping problems, systematic sequencing errors, and unusual allele balance, using the shell and perl scripts *RunVcfCombineUGHaplo_FDA_131103.sh, VcfCombineUGHaplo_v0.3.pl, VcfHighConfUGHaploMulti_HomJoint_1.3_FDA.pl, GATK_VQSR_..._131103.sh, and runVariantRecal..._131103.pl*. The annotations used for systematic sequencing errors, alignment bias, mapping bias, and abnormal allele balance for homozygous and heterozygous genotypes are listed in Supplementary Table S4. For each genomic position, the VQSR Gaussian mixture model outputs a tranche ranging from 0 to 100, with higher numbers indicating it has more unusual characteristics, which may indicate bias. For example, a tranche of 99.9 means that 0.1% of positions have characteristics more extreme than this position.

*Arbitration between datasets with conflicting genotypes*

For some positions in the genome, datasets have conflicting genotypes. Our approach to arbitrating between conflicting datasets is summarized in Fig. 1 and Supplementary Fig. S1. We hypothesize that if a dataset has unusual annotations associated with bias at a particular genome site, it is less likely to be correct than a dataset with typical characteristics at that genome site. For each possible variant site, we first determine if at least two datasets confidently call the same genotype (PLdiff>20) and at least 5x more datasets confidently call this genotype than disagree (i.e., have PLdiff>20 for a different genotype). In addition, when combining all datasets the net PLdiff and PLdiff/DP must exceed the values in Table S5 for the specific genotype and class of variant. Also, if two datasets have a indel called by the HaplotypeCaller within 20 bps and do not call a variant at this position, then it declared uncertain. If these conditions are not met, then we use the arbitration process. We start filtering the most unusual sites (tranche > 99.5). We first filter possible systematic sequencing errors above this tranche because they are most likely to be biases. Next, we filter possible alignment problems above this tranche. The order of tranche filtering is 99.5, 99, 95, and 90. We filter decreasing tranches until meeting the conditions above for PLdiff and PLdiff/DP.

Some positions in the genome are difficult for all methods, so even if all datasets agree on the genotype there may be significant uncertainty. For example, if a region has one copy in the hg19/GRCh37 reference assembly but two copies in both alleles in NA12878, and one of the copies has a homozygous SNP, it would incorrectly appear as a heterozygous SNP in all datasets. To minimize incorrect genotype calls, we use the VQSR tranches for annotations associated with systematic sequencing errors, alignment problems, mapping problems, and atypical allele balance. For homozygous reference genotypes, we require that at least 2 datasets have an alignment tranche < 99. For heterozygous genotypes, we require that at least 3 datasets have a mapping tranche < 99, at least 2 datasets have a systematic sequencing error tranche < 95, at least 2 datasets have an alignment tranche < 95, at least 2 datasets have an mapping tranche < 95, and at least 2 datasets have an allele balance tranche < 95. For homozygous variant genotypes, we require that at least 3 datasets have a mapping tranche < 99, at least 2 datasets have an alignment tranche < 99, and at least 2 datasets have an allele balance tranche < 99. For sites not considered potential variants, we determine whether they are callable as homozygous reference by using the GATK CallableLoci walker, requiring that at least three datasets have a coverage greater than 5, excluding base quality scores less than 10, and requiring that the fraction of reads with mapping quality <10 is <10% (CallableLoci_...sh). In addition, we remove all regions with known tandem duplications not in the GRCh37 Reference Assembly, and we optionally have a bed file that removes all structural variants for NA12878 reported in dbVar (as of June 12, 2013), and/or long homopolymers and tandem repeats that do not have at least 5 reads covering them in one of the datasets with 7 bp mapped on either side (created with BedSimpleRepeatBamCov.pl). We depict regions as "callable" using bed files, which is created using the process described above using MakeBedFiles_v2.18_131103.sh, with results and uncertain categories in Supplementary Tables S2 and S3. All bases inside the bed file and not in the variant call file can be considered highly confident homozygous reference, and can be used to assess FP rates in any sequencing dataset.

*GCAT performance assessment of dataset*

To perform the comparisons in GCAT, the variants in the vcf files were first regularized using vcflib vcfallelicprimitives. For the whole genome comparisons, the variants were also subset with the bed file excluding dbVar structural variants. For the whole exome comparisons, the variants were subset with both the bed file excluding dbVar structural variants and the target exome bed file from the manufacturer (iontorrent TargetSeq_hg19 http://ioncommunity.lifetechnologies.com/docs/DOC-2817 and Illumina exome ftp://ftp.1000genomes.ebi.ac.uk/vol1/ftp/technical/reference/exome_pull_down_targets//20130108.exome.targets.bed). Receiver operating characteristic (ROC) curves were generated by sorting the variants by coverage or variant quality score and calculating true positive rate and false positive rate as variants with decreasing coverage or variant quality score are added.

**Supplementary Information for Zook et al.**

**Manual inspection of alignments at discordant variants on chromosome 1 between our integrated calls and the X Prize fosmid calls**

To understand which method was correct when our integrated calls and the fosmid calls were discordant, we manually curated alignments from several datasets in the regions around a randomly selected 25% of the discordant variants (see Supplementary Figs. S4 to S28 for some examples). Manual curation of alignments from multiple datasets, aligners, and sequencing platforms allowed us to resolve the reasons for all of the differences between our integrated calls and the fosmid calls.

For the manually curated variants in our integrated calls and not in the fosmid calls, almost all were FNs in the fosmid calls due to mis-called complex variants (see Supplementary Fig. S4) or overly stringent filtering (see Supplementary Fig. S5), except for one SNP (chr2:108078636) that was clearly homozygous variant in all our integrated datasets but clearly homozygous reference in the fosmids (see Supplementary Fig. S6). The reason is unclear for the discordant SNP called from WGS but not in the fosmid reads, but it appears unlikely to be a FP in our integrated calls since coverage, alignments, mapping quality, base qualities, and other characteristics are all normal (see Supplementary Fig. S6). Also, one region that we only partially excluded as uncertain contained some highly confident variants that may actually be errors due to a large deletion (see Supplementary Fig. S7). In addition, our highly confident calls contained only part of a very large complex variant on chromosome 8, part of which was in our uncertain regions, though even more of the complex variant was missed in the fosmid calls (see Supplementary Fig. S8).

We also manually curated alignments around variants in the fosmid calls and not in our integrated calls, and several reasons were found for the differences. 19 of the discordant calls were compound heterozygous calls that were actually consistent, but the comparison algorithm does not appropriately compare them to haploid calls (see Supplementary Figs. S9 and S10). A few discordant calls resulted from systematic sequencing errors in the fosmid calls (see Supplementary Fig. S11). In addition, many of the discordant calls were errors in the fosmid complex calls (see Supplementary Figs. S4 and S5), but there were also 3 instances where our integrated calls call part of a complex variant uncertain so that a proper comparison could not be performed (see Supplementary Figs. S12 and S13). In addition, in one location with two nearby SNPs, our highly confident genotypes only contain one of them due to an error in our integration of UnifiedGenotyper and HaplotypeCaller calls (see Supplementary Fig. S14). There were also 8 SNPs and 7 indels for which the fosmid contained a clear variant and all of the WGS datasets contained strong evidence for a homozygous reference call without any unusual characteristics or evidence of bias (see Supplementary Figs. S15 and S16). Six of these indels were in homopolymers, and one was in a dinucleotide tandem repeat. These variants appear unlikely to be FNs in our integrated calls, but rather may result from low frequency de novo mutations in the cell line. The whole genome mutation rate of cell lines per doubling has not been well-measured and the number of doublings in this >30-year-old cell line is not known, but in a mutation rate of $1 \times 10^{-9}$ per generation over 3xx generations would generate ~15 mutations that would be picked up in fosmids covering 30 million bases. In one location, the fosmids called a SNP one base away from the correct location, likely due to a bug in the caller (see Supplementary Fig. S17). In several locations, the fosmids call a large indel of uncertain length, which is called correctly in our highly confident calls (see Supplementary Fig. S17).

The fosmids contain 119 million reference bases in our highly confident regions, and we manually curated 25% of discordant variants in these bases. Therefore, this analysis suggests that our integrated calls likely contain ~3 partial complex variant calls and between 0 and 1 false positive or false negative simple SNP or indel calls per 30 million highly confident bases, in which our integrated calls contain ~94,500 TP SNPs and ~1400 TP indels.

Table S1: Effects of limitations of the dataset used as a benchmark on performance assessment

| Benchmark dataset call | Assessed dataset is correct | | Assessed dataset is incorrect | |
|---|---|---|---|---|
| | Effect on False Positive Rate | Effect on False Negative Rate | Effect on False Positive Rate | Effect on False Negative Rate |
| False Positive | − | ↑ | ↓ | − |
| False Negative | ↑ | − | − | ↓ |
| Uncertain[1] | ↑ | ↑ | ↓ | ↓ |

[1]Uncertain calls will usually have a net effect of underestimating FP and FN rates because they usually disproportionately fall in more difficult regions of the genome

Table S2: Variants and regions included in the bed file describing highly confident regions as additional uncertain regions are excluded, with the percentage of total variants or bases in parentheses. Variants are from our integrated callset and from the 250bp whole genome Illumina called with GATK HaplotypeCaller v.2.6 (250bp_HC). Only bases that are not N in the reference genome and in chromosomes 1-22 and X are included.

| | Integrated | 250bp_HC Variants | | Non-N bases |
|---|---|---|---|---|
| | All | All | PASSonly | remaining in genome |
| Original variants | 10,819,577 (100%) | 6,390,200 (100%) | 6,222,108 (100%) | 2,835,690,481 (100%) |
| Low coverage or mapping quality | 6,291,061 (58.1%) | 6,135,423 (96.0%) | 6,020,657 (96.7%) | 2,752,862,066 (97.1%) |
| Add certain variants | 5,521,641 (51%) | 6,161,083 (96.4%) | 6,043,192 (97.1%) | 2,752,995,088 (97.1%) |
| Remove uncertain variants | 3,653,364 (33.8%) | 5,031,438 (78.7%) | 4,991,288 (80.2%) | 2,673,122,703 (94.3%) |
| Remove simple repeats | 3,620,149 (33.5%) | 4,976,032 (77.9%) | 4,936,900 (79.3%) | 2,662,216,037 (93.9%) |
| Remove segmental duplications | 3,495,411 (32.3%) | 4,794,353 (75.0%) | 4,771,193 (76.7%) | 2,586,928,427 (91.2%) |
| Remove decoy sequence | 3,493,963 (32.3%) | 4,792,819 (75.0%) | 4,770,405 (76.7%) | 2,586,546,317 (91.2%) |
| Remove RepeatSeq STRs | 3,339,354 (30.9%) | 4,571,495 (71.5%) | 4,550,470 (73.1%) | 2,484,884,293 (87.6%) |
| Remove structural variants | 2,915,732 (26.9%) | 4,001,758 (62.6%) | 3,987,453 (64.1%) | 2,195,078,292 (77.4%) |

Table S3: Variants and regions excluded as uncertain for different reasons during our integration process, and the numbers of variants that fall inside these regions from our integrated callset and from the 250bp whole genome Illumina called with GATK HaplotypeCaller v.2.6 (250bp_HC). Only bases that are not N in the reference genome and in chromosomes 1-22 and X are included.

|  | Bases Excluded | Integrated Variants All | Integrated Variants No filtered sites | 250bp_HC Variants All | 250bp_HC Variants No filtered sites |
|---|---|---|---|---|---|
| Mapping bias | 50,115 | 50,115 | - | 39,047 | 27,564 |
| Systematic sequencing error | 10,888 | 10,888 | - | 7,104 | 6,268 |
| Abnormal allele balance | 114,447 | 114,447 | - | 25,250 | 21,272 |
| Local Alignment bias | 87,763 | 87,763 | - | 14,337 | 11,559 |
| < 2 datasets | 59,195 | 59,195 | - | 7,505 | 7,001 |
| Low coverage | 34,436 | 34,436 | - | 1,812 | 1,688 |
| Reference in HaplotypeCaller | 13,941 | 13,941 | - | 5,501 | 5,259 |
| Conflicting genotypes | 1,443,677 | 1,443,680 | - | 342,945 | 265,930 |
| Low coverage/low mapping quality | 82,828,415 | 795,436 | 60,580 | 254,777 | 201,451 |
| Segmental duplications | 150,638,985 | 781,808 | 165,801 | 414,780 | 323,818 |
| 1000 Genomes decoy | 1,507,000 | 45,166 | 16,165 | 7,376 | 3,603 |
| Simple Repeats | 18,651,604 | 351,622 | 76,421 | 227,238 | 220,740 |
| dbVar Structural Variants | 432,456,384 | 1,580,710 | 595,655 | 1,074,436 | 980,695 |

Supplementary Table S4: Annotations used in GATK Variant Quality Score Recalibration for arbitration and flagging difficult sites as uncertain

| Category of Bias | Homozygous calls | Heterozygous calls |
|---|---|---|
| Systematic sequencing errors | Neighboring base quality score* | Fisher Strand Bias (FS)<br>Base Quality Rank Sum Test (BaseQRankSum)<br>Neighboring base quality score* |
| Alignment bias | Mean distance from either end of the read (ReadPosEndDist)* | HaplotypeScore<br>Mean distance from either end of the read (ReadPosEndDist)* |
| Mapping bias[1] | Mean Mapping Quality<br>Fraction of reads with MQ=0 (MQ0Fraction)<br>Mapping Quality Rank Sum Test (MQRankSum)<br>Depth of Coverage (DP) | Mean Mapping Quality<br>Fraction of reads with MQ=0 (MQ0Fraction)<br>Mapping Quality Rank Sum Test (MQRankSum)<br>Depth of Coverage (DP) |
| Abnormal allele balance[2] | Allele Balance<br>Variant Quality Score/Depth of Coverage (QD) | Allele Balance<br>Variant Quality Score/Depth of Coverage (QD) |

[1]Mapping bias is only used for flagging heterozygous sites as uncertain and not for arbitrating between datasets
[2]Abnormal allele balance is only used for flagging heterozygous and homozygous variant sites as uncertain
*Annotations for GATK developed in this work and available as part of the bcbio.variation package

Supplementary Table S5: Cutoffs for the most likely genotype likelihood ratio (PLdiff) and PLdiff divided by depth of coverage (PLdiff/DP), which were used for determining whether a genotype is confident or uncertain.

| Likelihood ratio cutoff | Homozygous SNPs | Heterozygous SNPs | Homozygous indels | Heterozygous indels |
|---|---|---|---|---|
| **PLdiff** | 120 | 200 | 80 | 100 |
| **PLdiff/DP** | 1.6 | 6.8 | 0.8 | 3.4 |

| | | | |
|---|---|---|---|
| **Input:** | Mapped reads (bam files) from each dataset | | |
| **Find all possible variant sites for each bam file:** | Calls SNPs and indels with GATK UnifiedGenotyper (QUAL>=2) *GATK_..._UG_...sh* | Calls SNPs and indels with GATK HaplotypeCaller (QUAL>=2) *GATK_..._haplo_array...sh* | Calls SNPs and indels with Cortex *de novo* assembly for Illumina PCR-free dataset *Run_cortex...sh* |
| **Find union of vcfs:** | Find union of SNP and indel calls from each variant caller from each dataset (*CombineVariants....sh*) | | |
| **Force calls in each bam file using union sites:** | Force GATK UnifiedGenotyper (UG) to genotype union SNP/indel sites *GATK_..._UG_recall...sh* | Force GATK HaplotypeCaller (HC) to perform local *de novo* assembly around union indel sites *GATK_haplo_recall_...sh* | |
| **Combine UG and HC calls for each bam file:** | Combine calls from UG and HC, using calls from HC plus the calls from UG that are not within 20bp of a confident HC indel call *RunVcfCombineUGHaplo...sh (VcfCombineUGHaplo...pl)* | | |
| **Find concordant genotype calls from combined vcfs to train VQSR:** | Find highly confident calls to train VQSR, requiring high genotype likelihood (PL), high PL/coverage (surrogate for allele balance), few datasets with a different genotype, and few datasets with poor mapping quality for het calls (see methods for details). *RunVcfCombineUGHaplo...sh (VcfHighConfUGHaploMulti...pl)* | | |
| **Find evidence of systematic errors in each bam file:** | Use GATK VQSR to find evidence of systematic sequencing, alignment, and mapping errors and atypical allele balance separately for homozygous and heterozygous SNPs and indels *GATK_VQSR...sh* | | |
| **Integrate datasets to form highly confident calls:** | Where datasets have discordant genotype calls, filter datasets with evidence of bias, decreasing threshold for bias until at least 5x more datasets agree than disagree, and there is a high combined PL and PL/coverage. Also mark sites as uncertain if fewer than 2 datasets confidently call the genotype without evidence of systematic error *VcfClassifyUsingVQSR...sh* | | |

Fig. S1: Detailed process for integrating genotype calls from multiple sequencing datasets by using evidence of bias to arbitrate between discordant datasets. Italics indicate scripts responsible for each step.

Fig. S2: Example of arbitration using characteristics of alignment bias. In this case, one allele has a G>A SNP followed by a TCCG insertion 8 bases downstream. Bwa with GATK Indel Realignment properly aligns longer Illumina HiSeq reads in this region (top), but bwa alone does not properly align shorter Illumina GAIIx reads in this 4-bp repeat region (bottom). Our arbitration process ignores GAIIx results at this position because it has characteristics of alignment bias due to clipping of aligned reads, including short aligned reads and bases falling near the end of aligned reads.

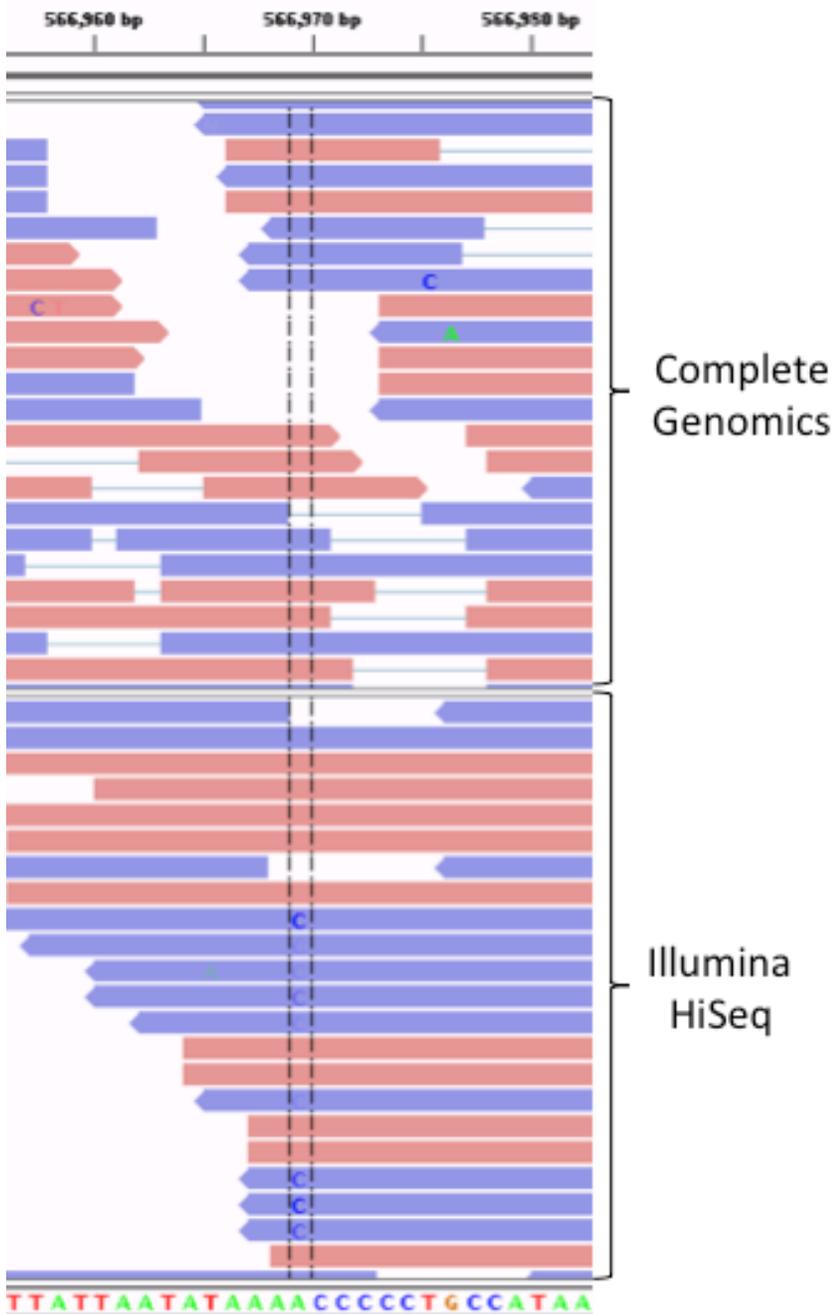

Fig. S3: Example of arbitration at position 566,969 on chromosome 1 using evidence of systematic sequencing errors. In this case, Illumina HiSeq has a systematic A>C error only on the reverse strand (blue) due to the G homopolymer followed by a T (A followed by C homopolymer on the forward strand). Complete Genomics does not have strand bias at this position, so we use it along with other datasets to call this location homozygous reference.

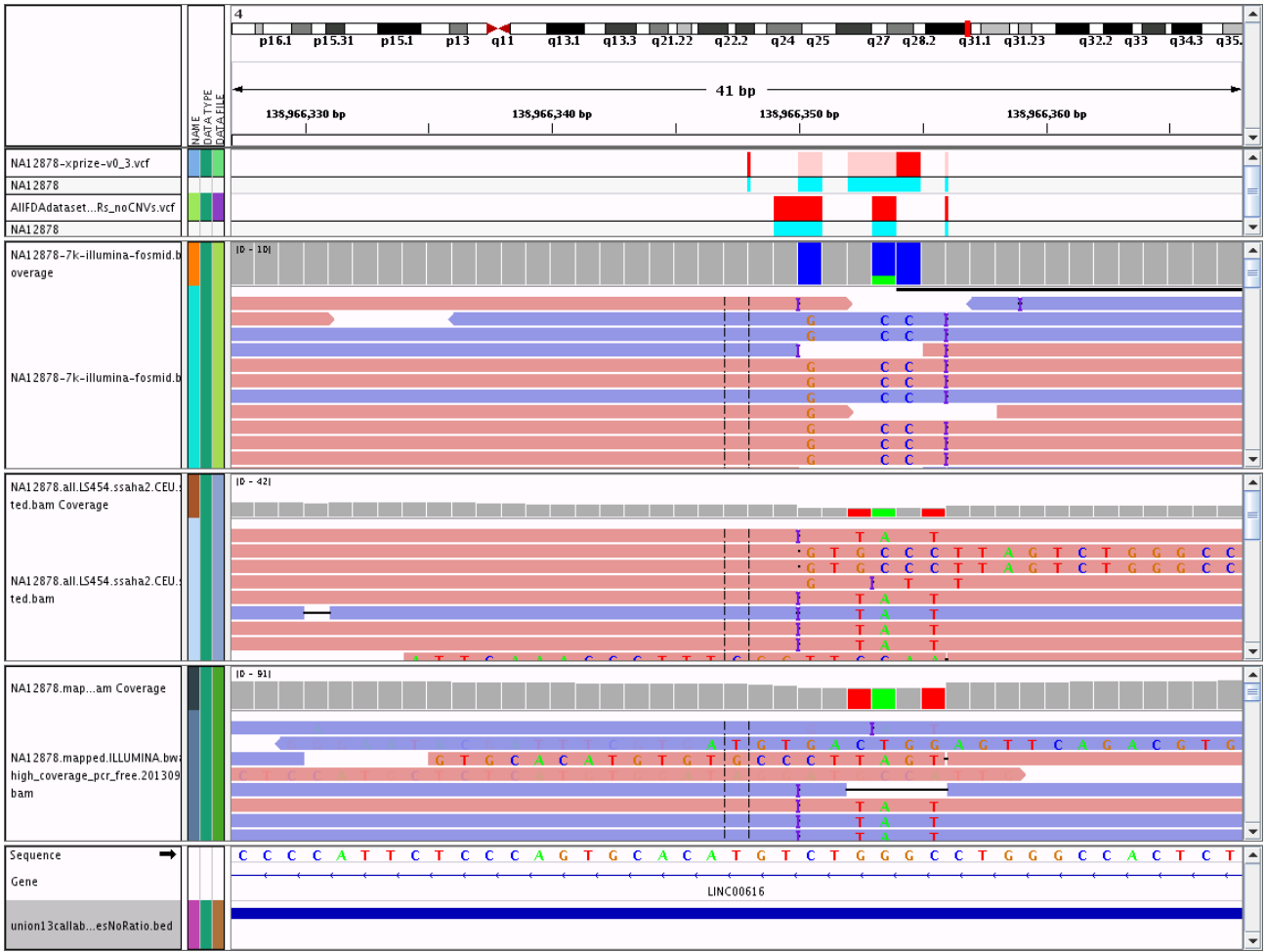

Fig. S4: Example of complex variant that is only partially called in the fosmid vcf and fully called correctly in our highly confident genotypes. Displayed from top to bottom are the fosmid vcf, our highly confident vcf, the fosmid alignments, 454 whole genome alignments, 250x250 bp Illumina alignments, and our highly confident regions.

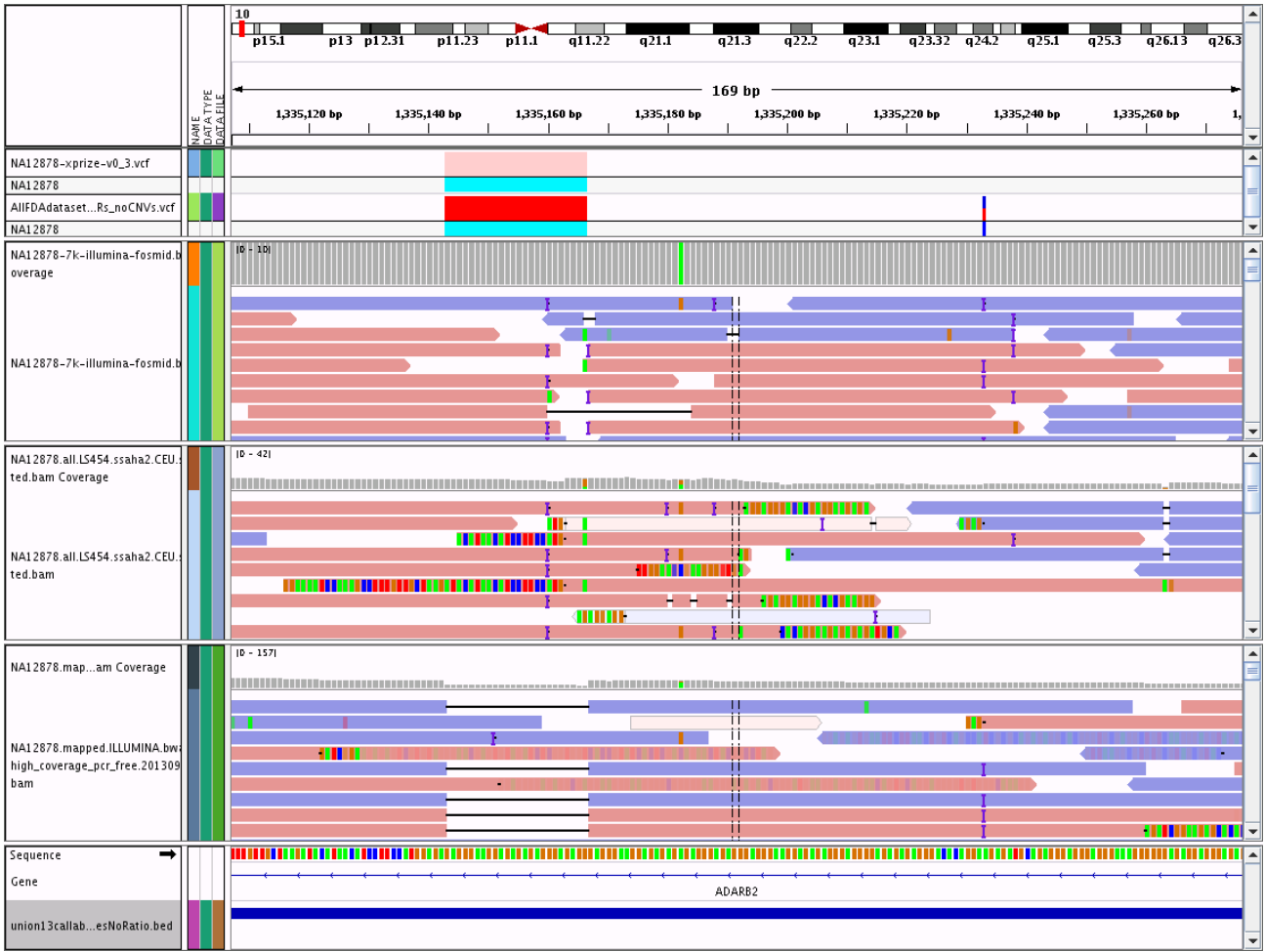

Fig. S5: Example of complex variant that is partially filtered and partially missed in the fosmid vcf and fully called in our highly confident genotypes. Displayed from top to bottom are the fosmid vcf, our highly confident vcf, the fosmid alignments, 454 whole genome alignments, 250x250 bp Illumina alignments, and our highly confident regions.

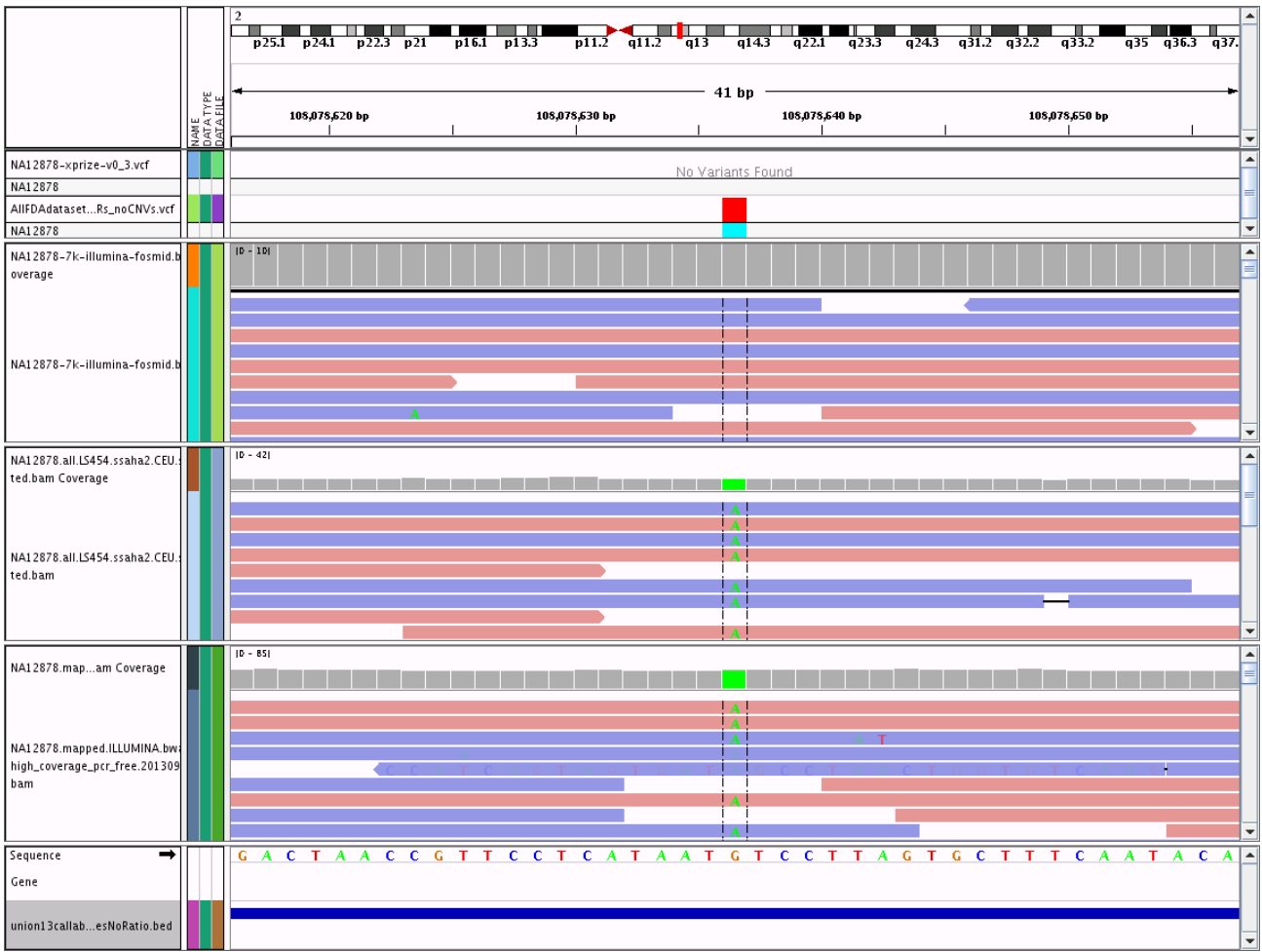

Fig. S6: Example of homozygous variant that is clearly homozygous reference in the fosmid vcf and clearly homozygous variant in the datasets used to form our highly confident genotypes. Displayed from top to bottom are the fosmid vcf, our highly confident vcf, the fosmid alignments, 454 whole genome alignments, 250x250 bp Illumina alignments, and our highly confident regions.

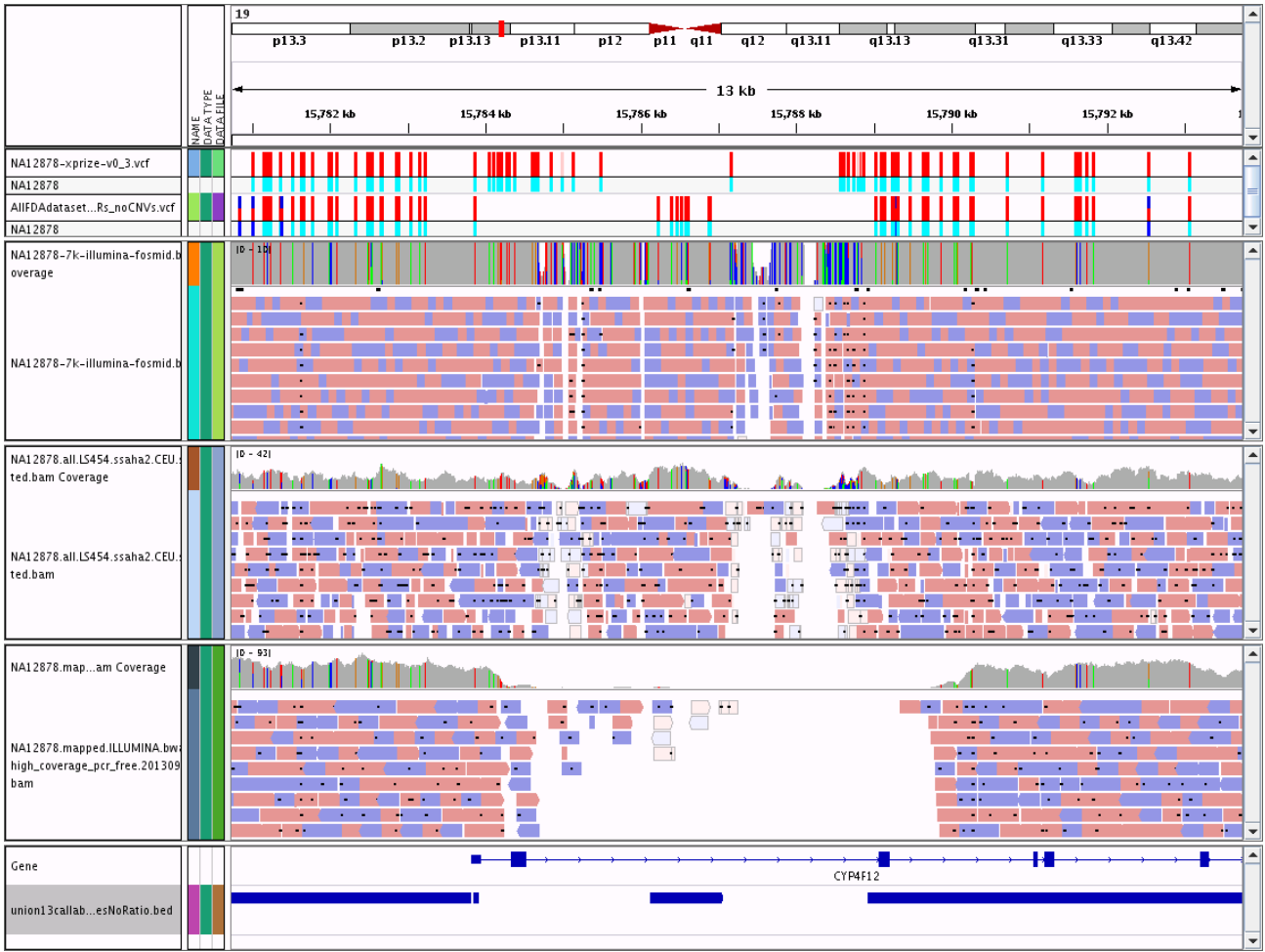

Fig. S7: Example of a region that may be a homozygous deletion and is mostly but not entirely excluded from our highly confident regions. Displayed from top to bottom are the fosmid vcf, our highly confident vcf, the fosmid alignments, 454 whole genome alignments, 250x250 bp Illumina alignments, and our highly confident regions.

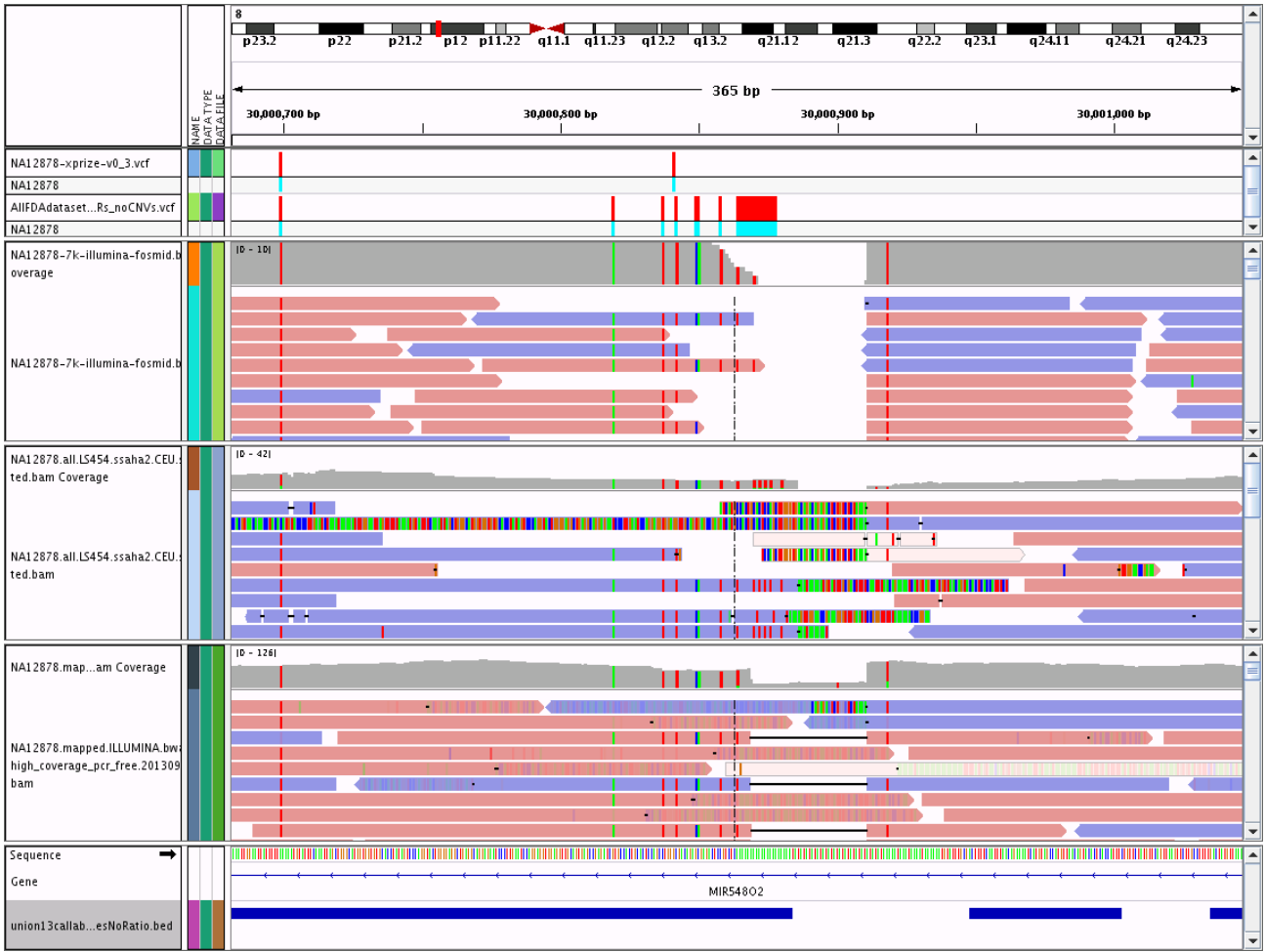

Fig. S8: Example of a likely large complex variant that is mostly missed in the fosmid vcf, and is partially called in our highly confident genotypes, but is partially excluded as uncertain from our highly confident regions. Displayed from top to bottom are the fosmid vcf, our highly confident vcf, the fosmid alignments, 454 whole genome alignments, 250x250 bp Illumina alignments, and our highly confident regions.

Fig. S9: Example of compound 4-bp/6-bp heterozygous deletion that is called correctly in the fosmid vcf and in our highly confident genotypes, but the comparison algorithm does not recognize as consistent. Displayed from top to bottom are the fosmid vcf, our highly confident vcf, the fosmid alignments, 454 whole genome alignments, 250x250 bp Illumina alignments, and our highly confident regions.

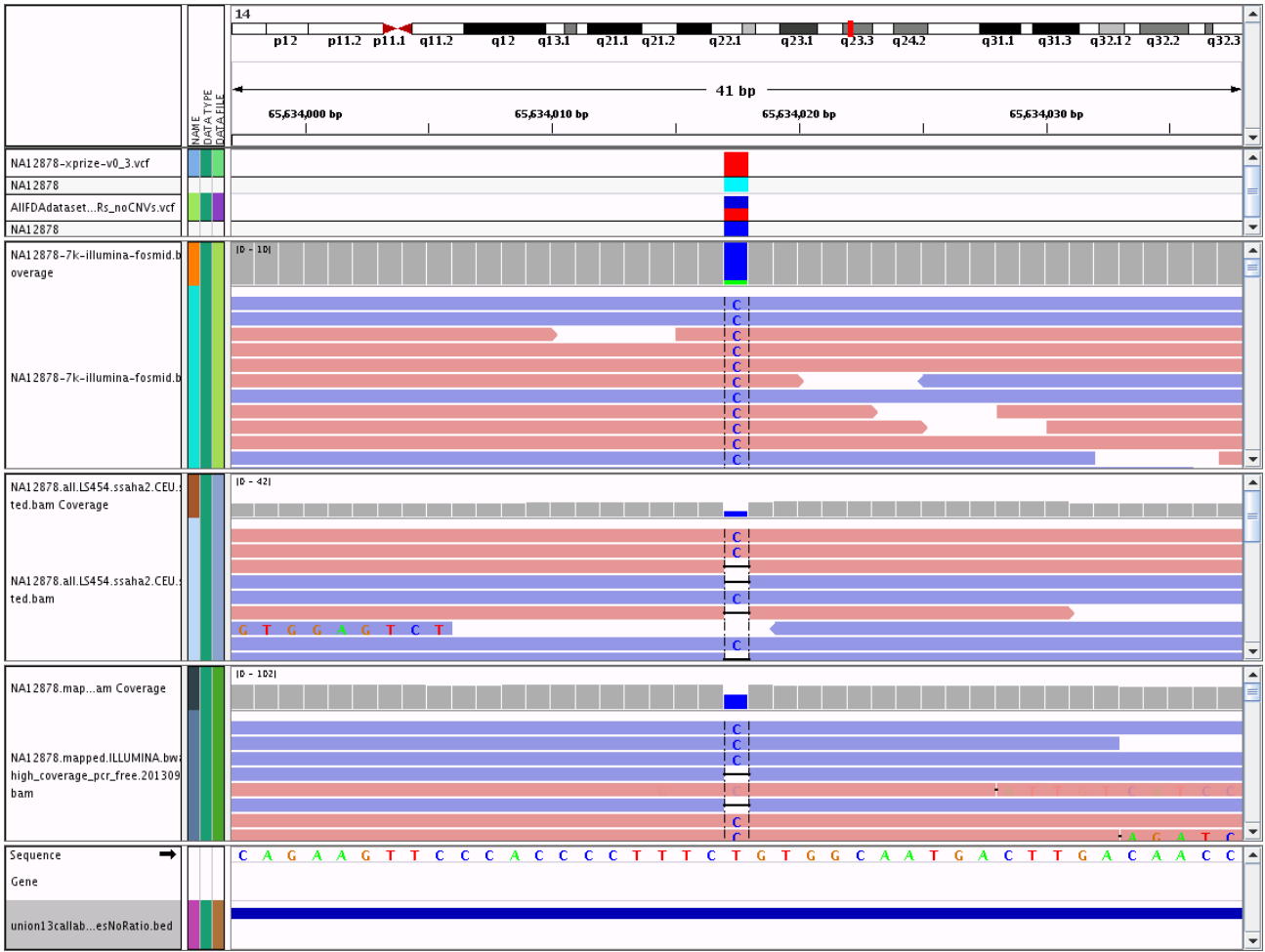

Fig. S10: Example of a compound SNP/deletion heterozygote, for which the SNP is called correctly in the fosmid vcf, and our highly confident vcf correctly calls a compound heterozygous SNP/deletion. Displayed from top to bottom are the fosmid vcf, our highly confident vcf, the fosmid alignments, 454 whole genome alignments, 250x250 bp Illumina alignments, and our highly confident regions.

Fig. S11: Example of a false positive 1-bp homopolymer expansion in the fosmid vcf (likely due to homopolymer sequencing or PCR errors), and is called correctly in our highly confident vcf, resulting in both a FP call in the fosmids. Displayed from top to bottom are the fosmid vcf, our highly confident vcf, the fosmid alignments, 454 whole genome alignments, 250x250 bp Illumina alignments, and our highly confident regions.

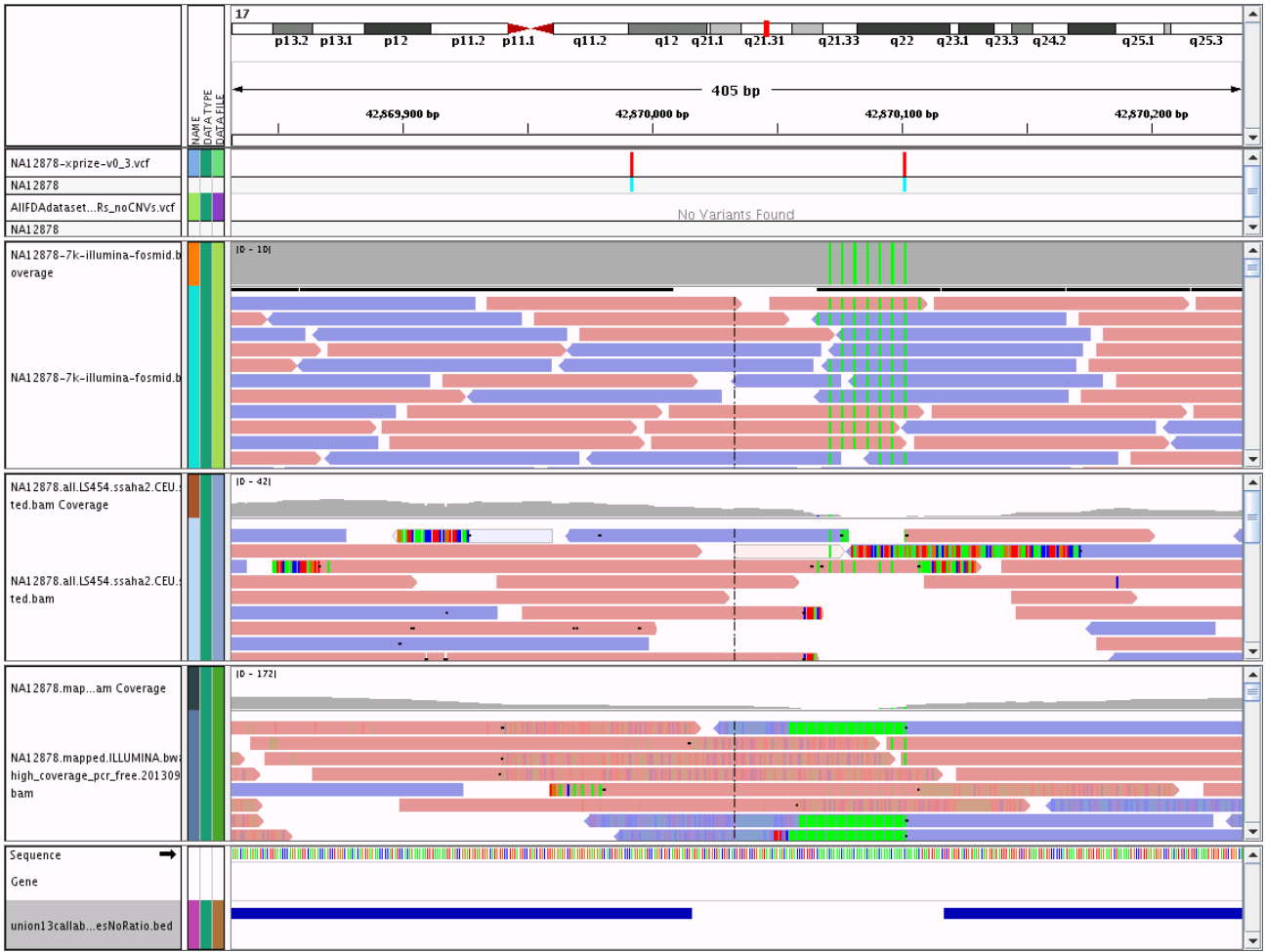

Fig. S12: Example of a possible very large deletion that is mis-called in the fosmid vcf, but is only partially excluded as uncertain in our highly confident vcf. Displayed from top to bottom are the fosmid vcf, our highly confident vcf, the fosmid alignments, 454 whole genome alignments, 250x250 bp Illumina alignments, and our highly confident regions.

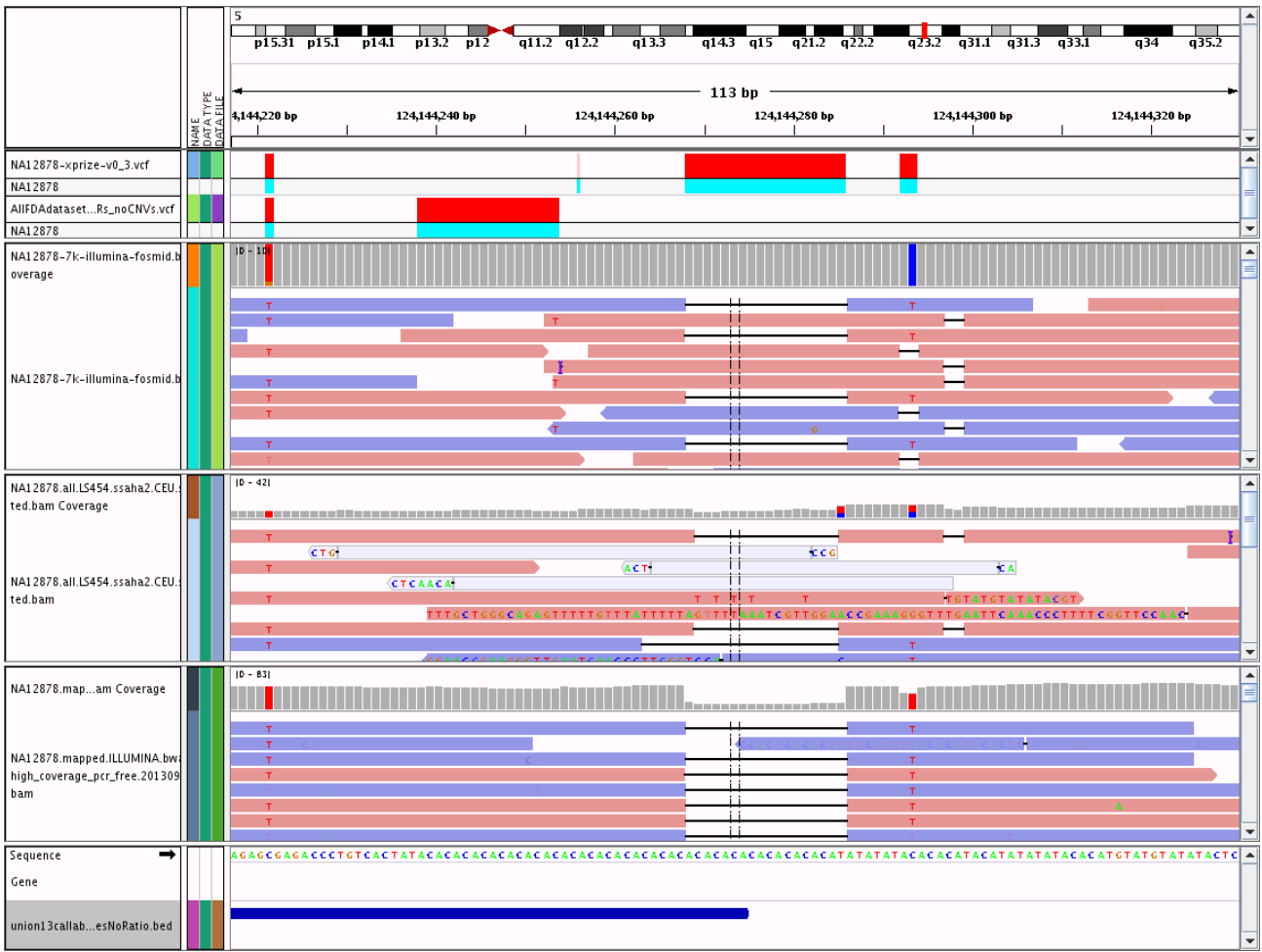

Fig. S13: Example of a very large complex variant that is called correctly in the fosmid vcf, and is partially called correctly in our highly confident vcf but is partially excluded by our highly confident regions. Displayed from top to bottom are the fosmid vcf, our highly confident vcf, the fosmid alignments, 454 whole genome alignments, 250x250 bp Illumina alignments, and our highly confident regions.

Fig. S14: Example of two nearby SNPs that are called correctly in the fosmid vcf, but only one is called in our highly confident vcf. Displayed from top to bottom are the fosmid vcf, our highly confident vcf, the fosmid alignments, 454 whole genome alignments, 250x250 bp Illumina alignments, and our highly confident regions.

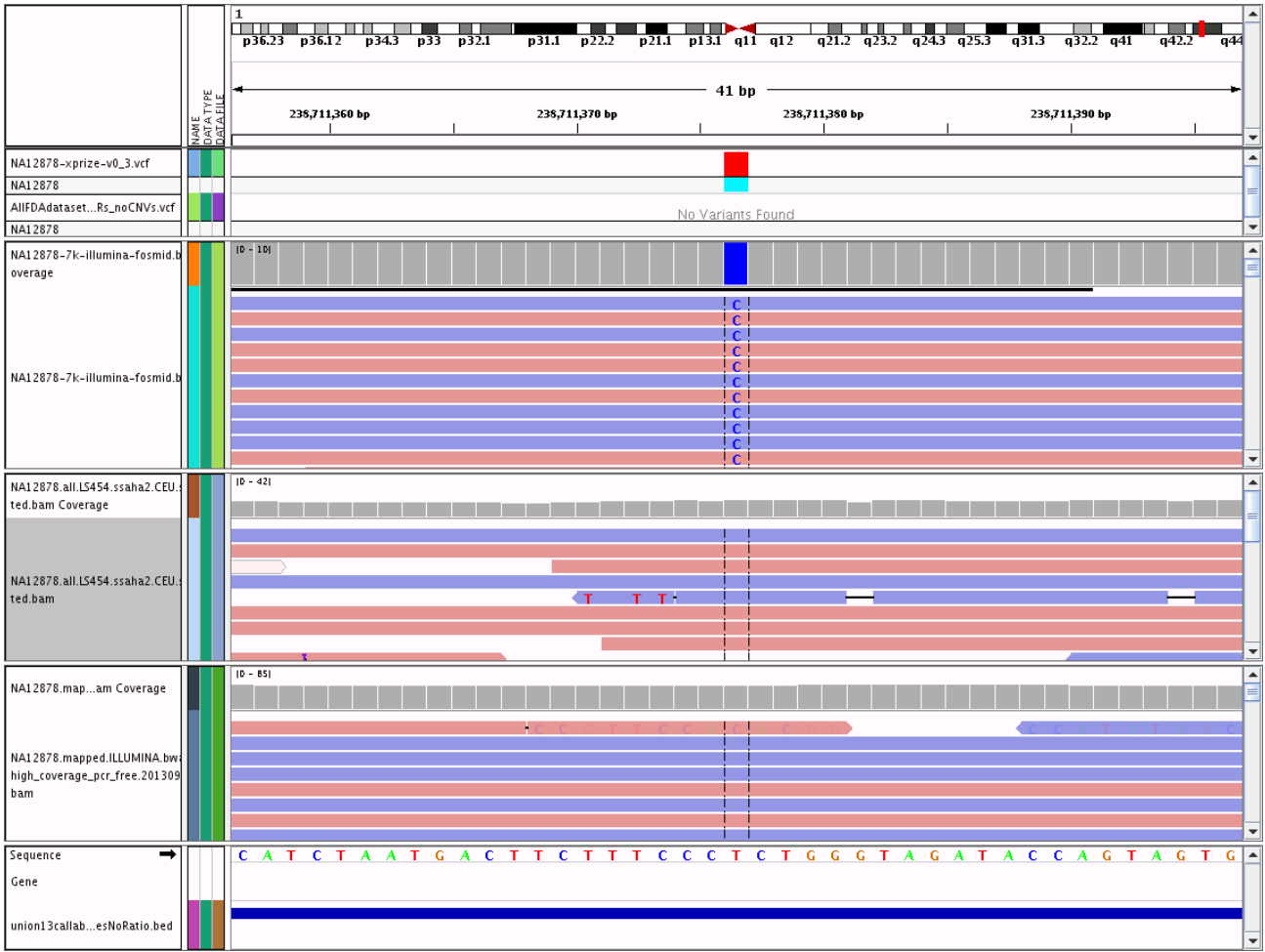

Fig. S15: Example of a variant that is clearly homozygous in the fosmid vcf but has no evidence in any of the whole genome or exome datasets (including those not shown), possibly due to a de novo mutation in the cell line or fosmid. Displayed from top to bottom are the fosmid vcf, our highly confident vcf, the fosmid alignments, 454 whole genome alignments, 250x250 bp Illumina alignments, and our highly confident regions.

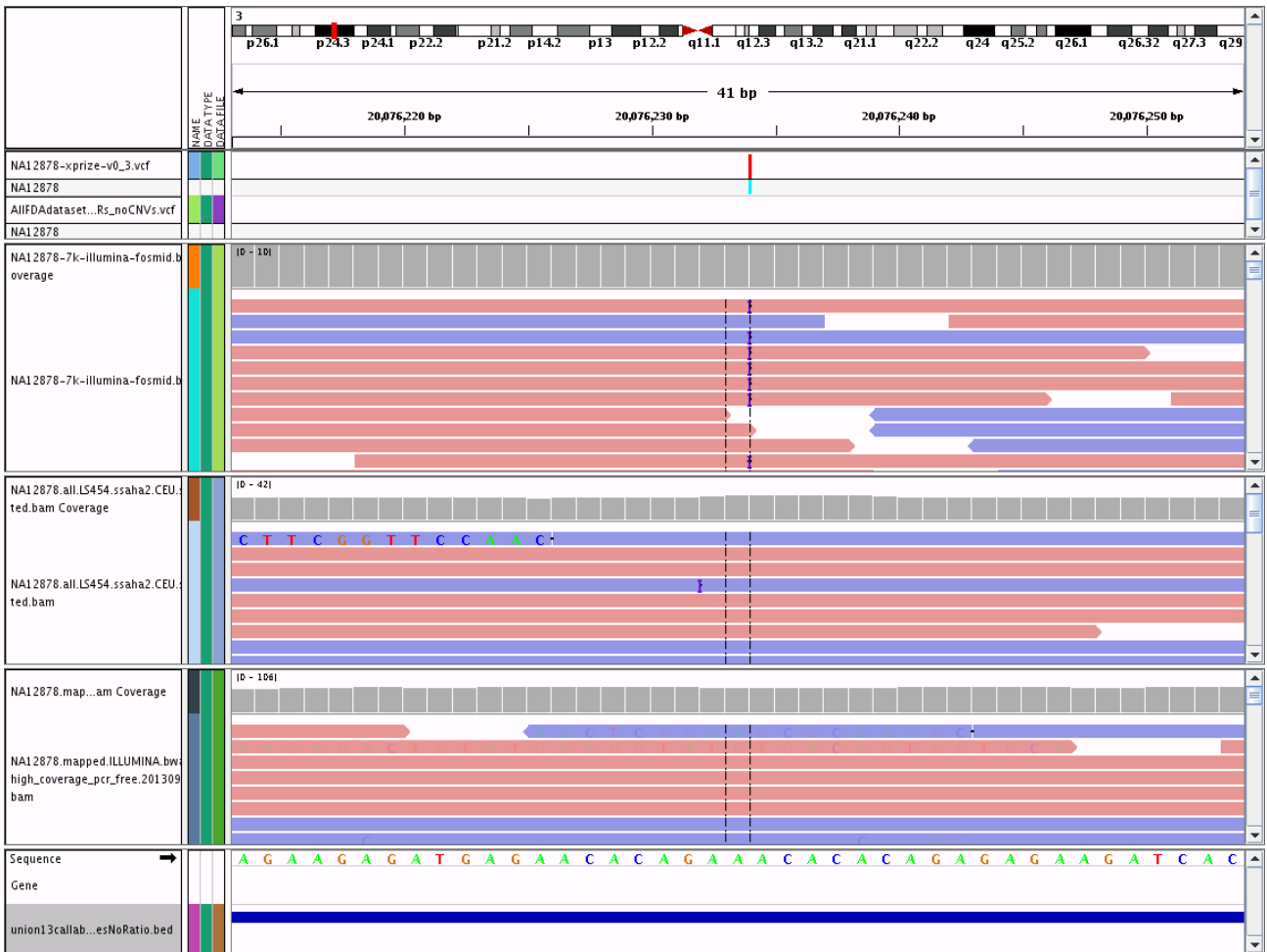

Fig. S16: Example of a variant (C insertion) that is clearly homozygous in the fosmid vcf but has no evidence in any of the whole genome or exome datasets (including those not shown), possibly due to a de novo mutation in the cell line or fosmid. Displayed from top to bottom are the fosmid vcf, our highly confident vcf, the fosmid alignments, 454 whole genome alignments, 250x250 bp Illumina alignments, and our highly confident regions.

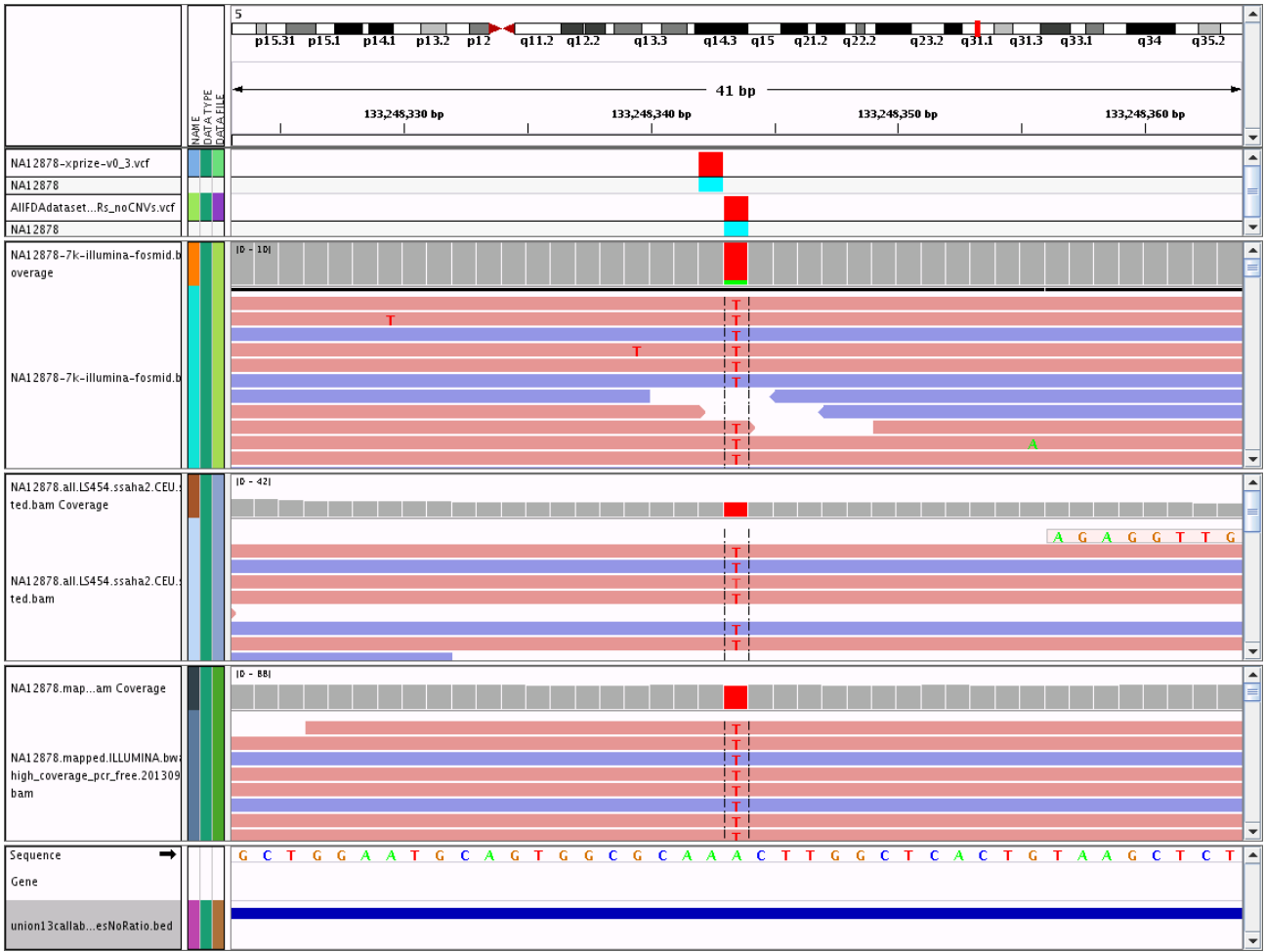

Fig. S17: Example of a variant that is called in the incorrect position in the fosmid vcf, and is called correctly in our highly confident vcf, resulting in both a FP and FN call in the fosmids. Displayed from top to bottom are the fosmid vcf, our highly confident vcf, the fosmid alignments, 454 whole genome alignments, 250x250 bp Illumina alignments, and our highly confident regions.

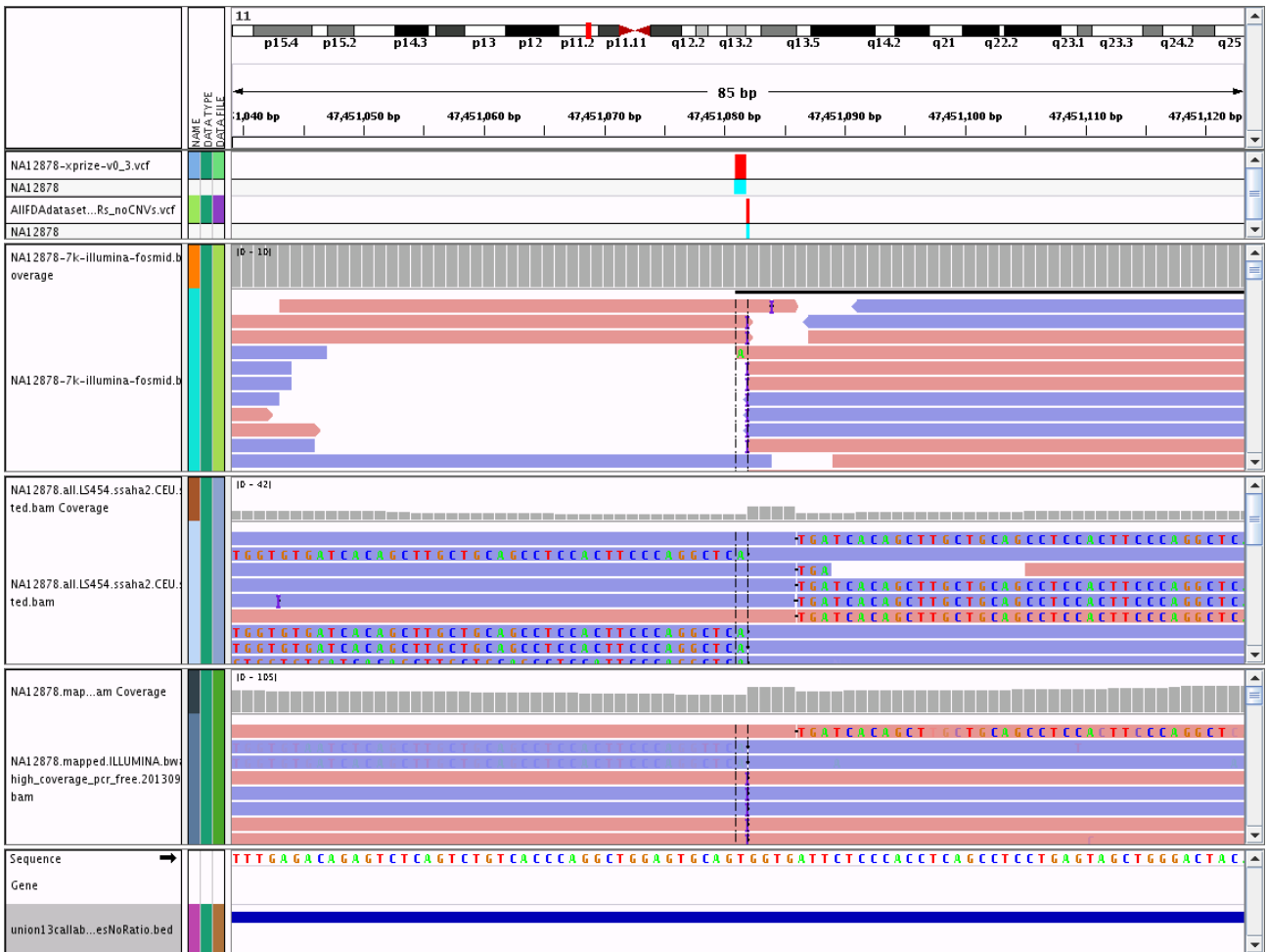

Fig. S18: Example of a large insertion that is called as an insertion of uncertain size or content in the fosmid vcf, and is called correctly in our highly confident vcf. Displayed from top to bottom are the fosmid vcf, our highly confident vcf, the fosmid alignments, 454 whole genome alignments, 250x250 bp Illumina alignments, and our highly confident regions.

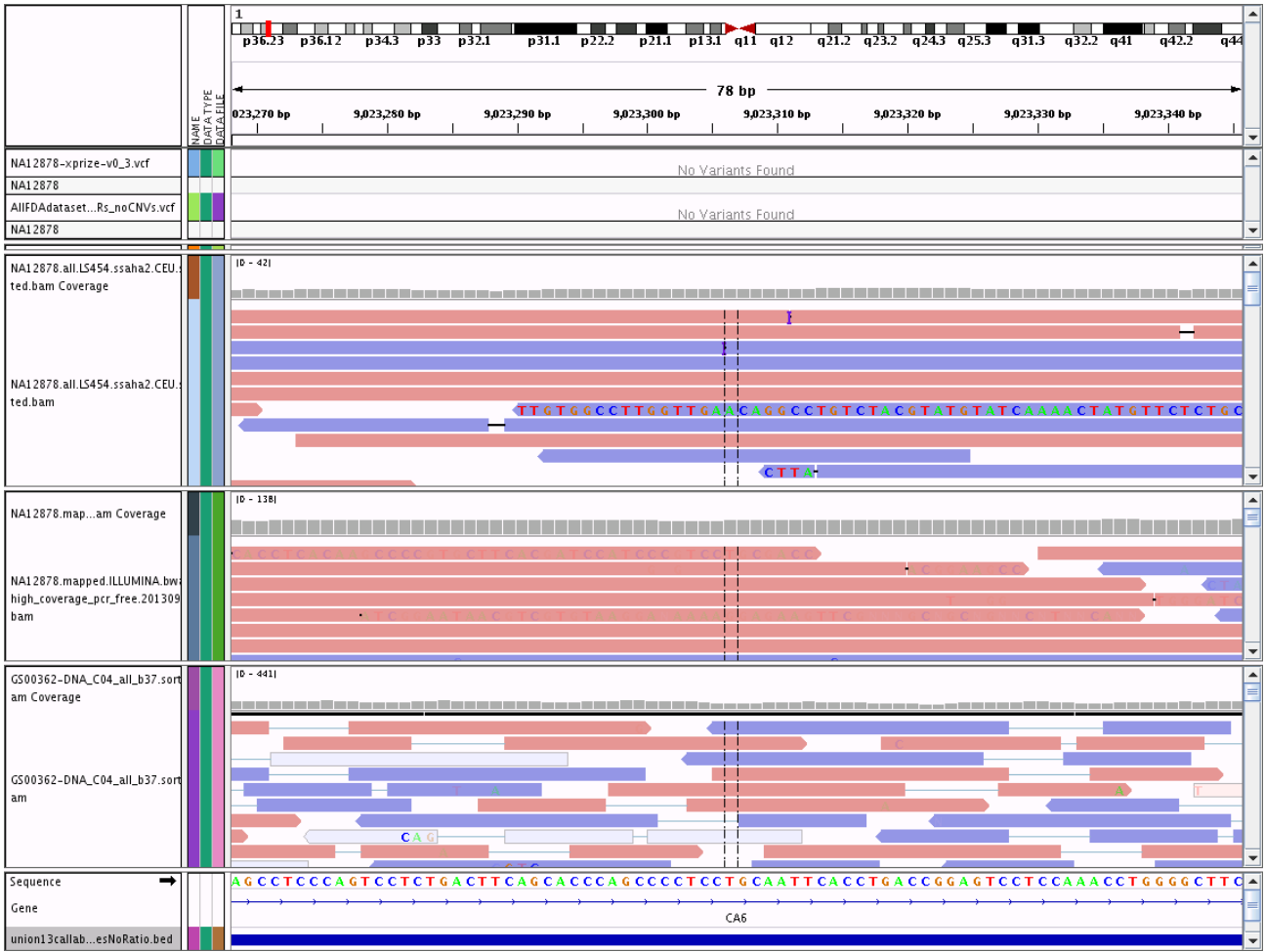

Fig. S19: Example of a location that is called a SNP by the OMNI microarray but all sequencing datasets clearly call homozygous reference. Displayed from top to bottom are the fosmid vcf, our highly confident vcf, 454 whole genome alignments, 250x250 bp Illumina alignments, Complete Genomics alignments, and our highly confident regions.

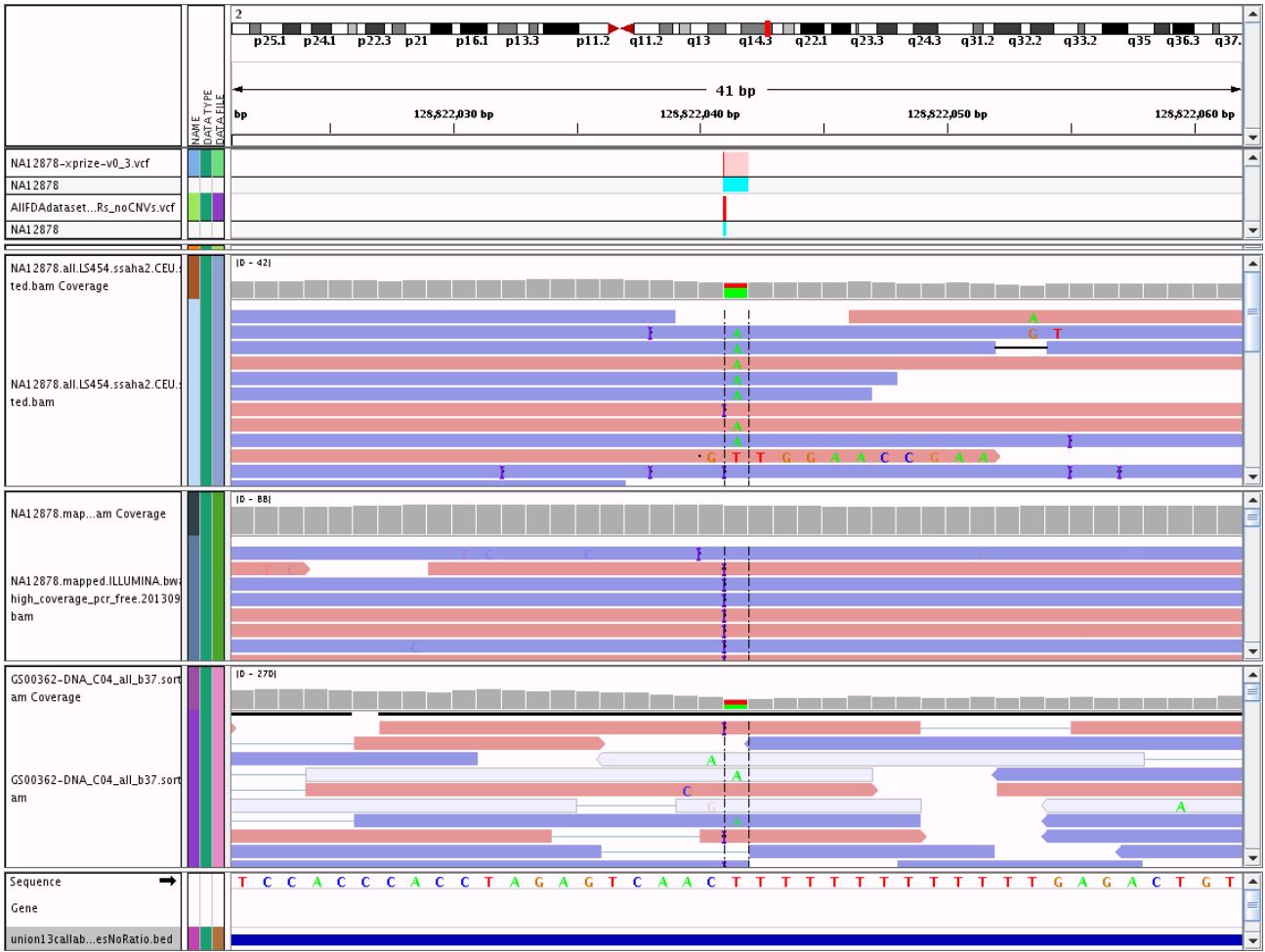

Fig. S20: Example of a location that is called a SNP by the OMNI microarray but appears more likely to be an A insertion before a T homopolymer. Displayed from top to bottom are the fosmid vcf, our highly confident vcf, 454 whole genome alignments, 250x250 bp Illumina alignments, Complete Genomics alignments, and our highly confident regions.

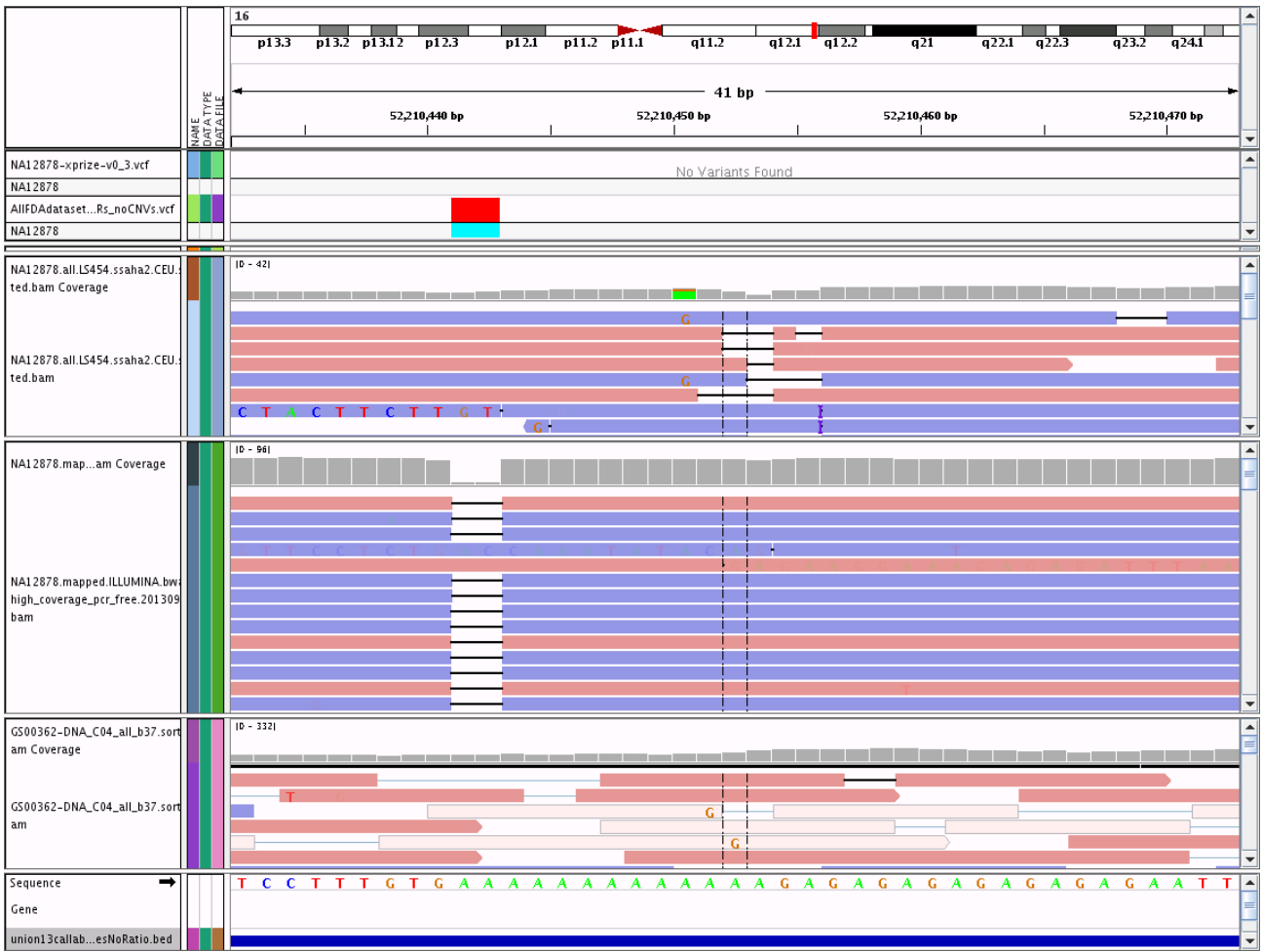

Fig. S21: Example of a location that is called a SNP by the OMNI microarray but appears more likely to be an AA deletion in an A homopolymer adjacent to an AG tandem repeat. Displayed from top to bottom are the fosmid vcf, our highly confident vcf, 454 whole genome alignments, 250x250 bp Illumina alignments, Complete Genomics alignments, and our highly confident regions.

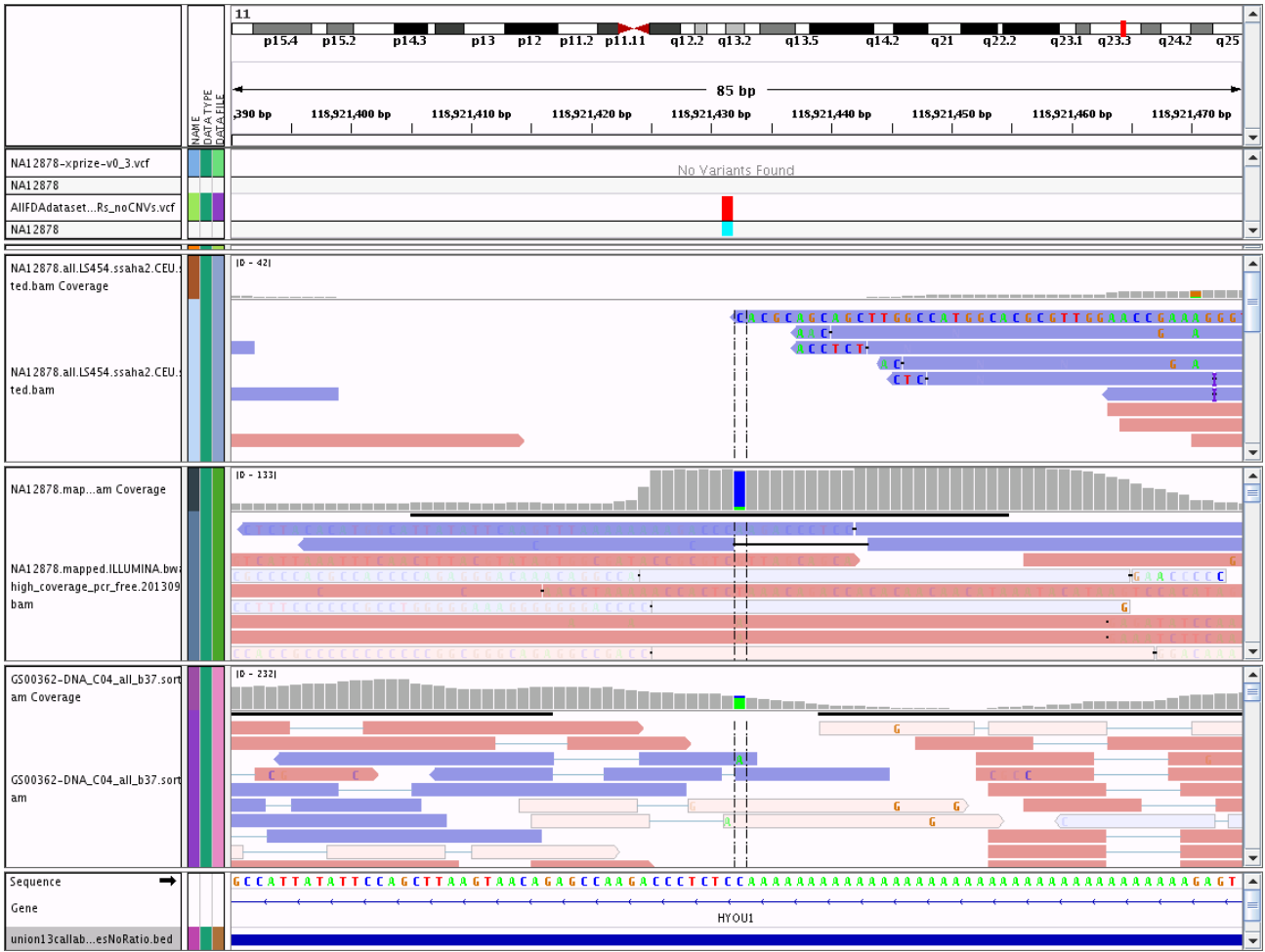

Fig. S22: Example of a location that is called a SNP by the OMNI microarray and is unclear in all sequencing datasets because no reads completely traverse the A homopolymer. Displayed from top to bottom are the fosmid vcf, our highly confident vcf, 454 whole genome alignments, 250x250 bp Illumina alignments, Complete Genomics alignments, and our highly confident regions.

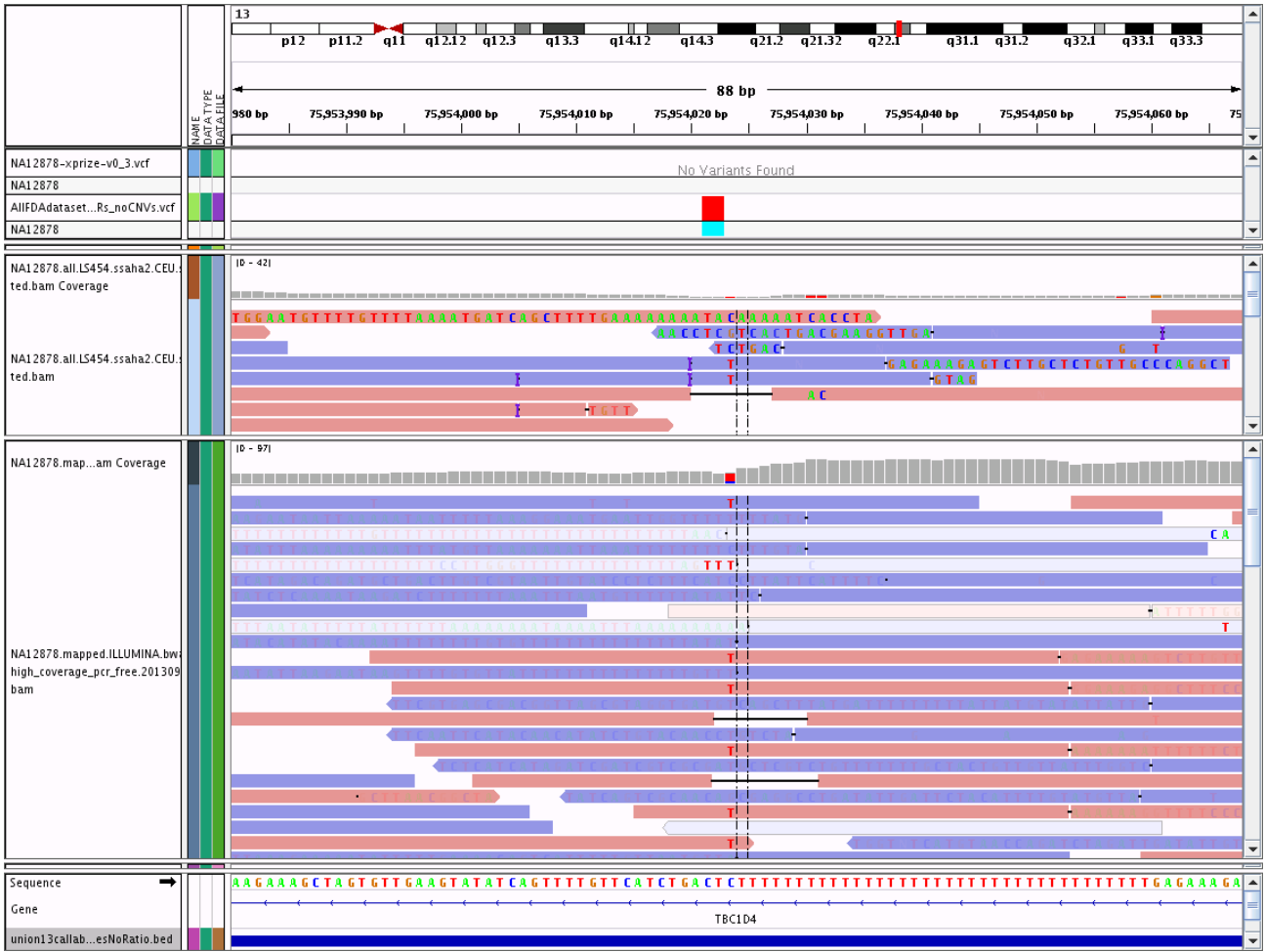

Fig. S23: Example of a location that is called a SNP by the OMNI microarray and a 2-bp deletion in our highly confident calls, but appears more likely to be an 8-bp deletion in a long T homopolymer based on Illumina PCR-free reads that completely traverse the homopolymer. Displayed from top to bottom are the fosmid vcf, our highly confident vcf, 454 whole genome alignments, 250x250 bp Illumina alignments, Complete Genomics alignments, and our highly confident regions.

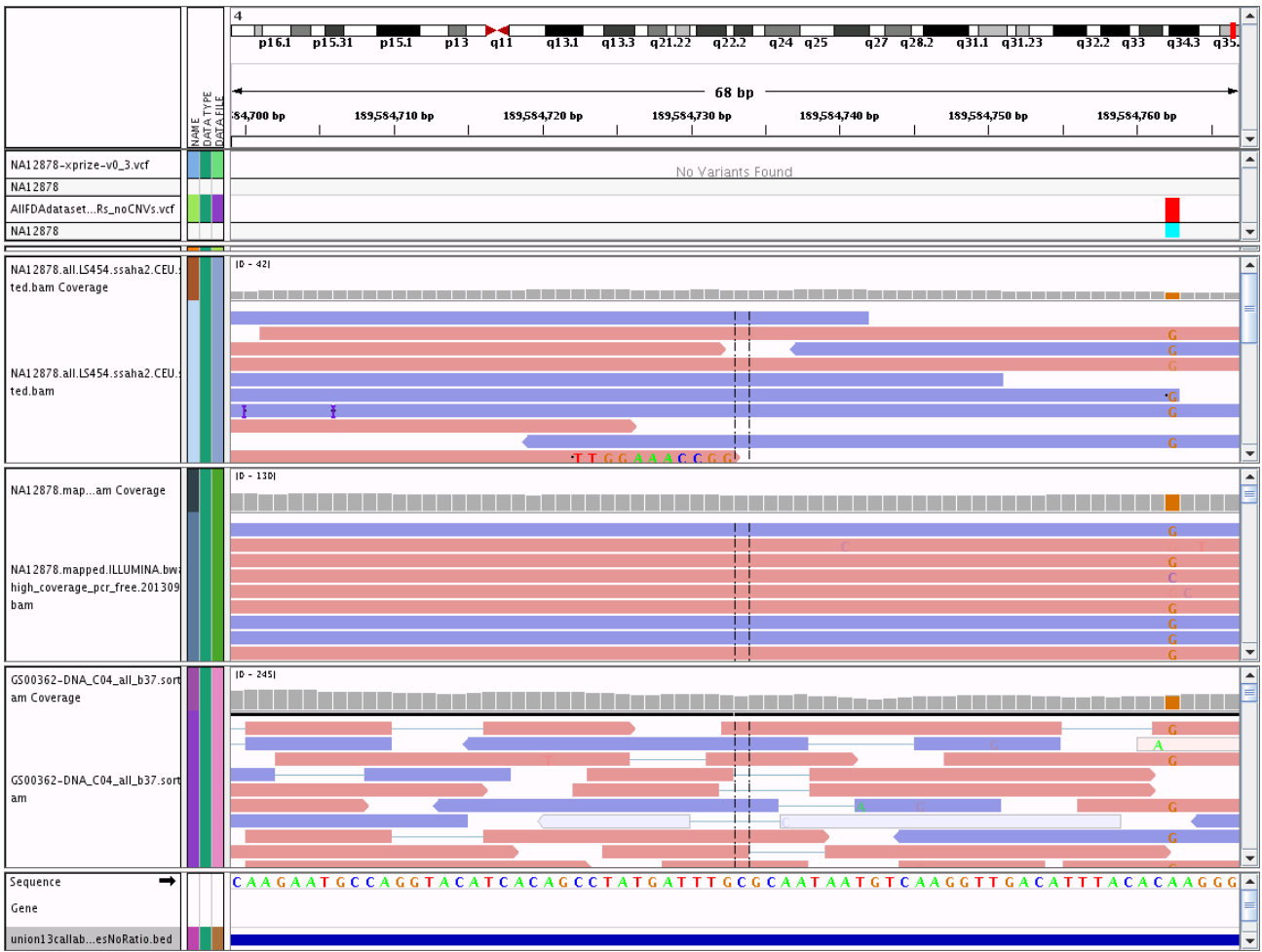

Fig. S24: Example of a location that is called a SNP by the OMNI microarray that has no evidence in sequencing, but might be confounded by a SNP downstream. Displayed from top to bottom are the fosmid vcf, our highly confident vcf, 454 whole genome alignments, 250x250 bp Illumina alignments, Complete Genomics alignments, and our highly confident regions.

Fig. S25: Example of a location that is called a SNP by the OMNI microarray but is actually a complex variant that can be represented in multiple ways. Displayed from top to bottom are the fosmid vcf, our highly confident vcf, 454 whole genome alignments, 250x250 bp Illumina alignments, Complete Genomics alignments, and our highly confident regions.

Fig. S26: Example of a location that is called homozygous reference by the OMNI microarray but is actually a homozygous deletion called correctly by our highly confident calls. Displayed from top to bottom are the fosmid vcf, our highly confident vcf, 454 whole genome alignments, 250x250 bp Illumina alignments, Complete Genomics alignments, and our highly confident regions.

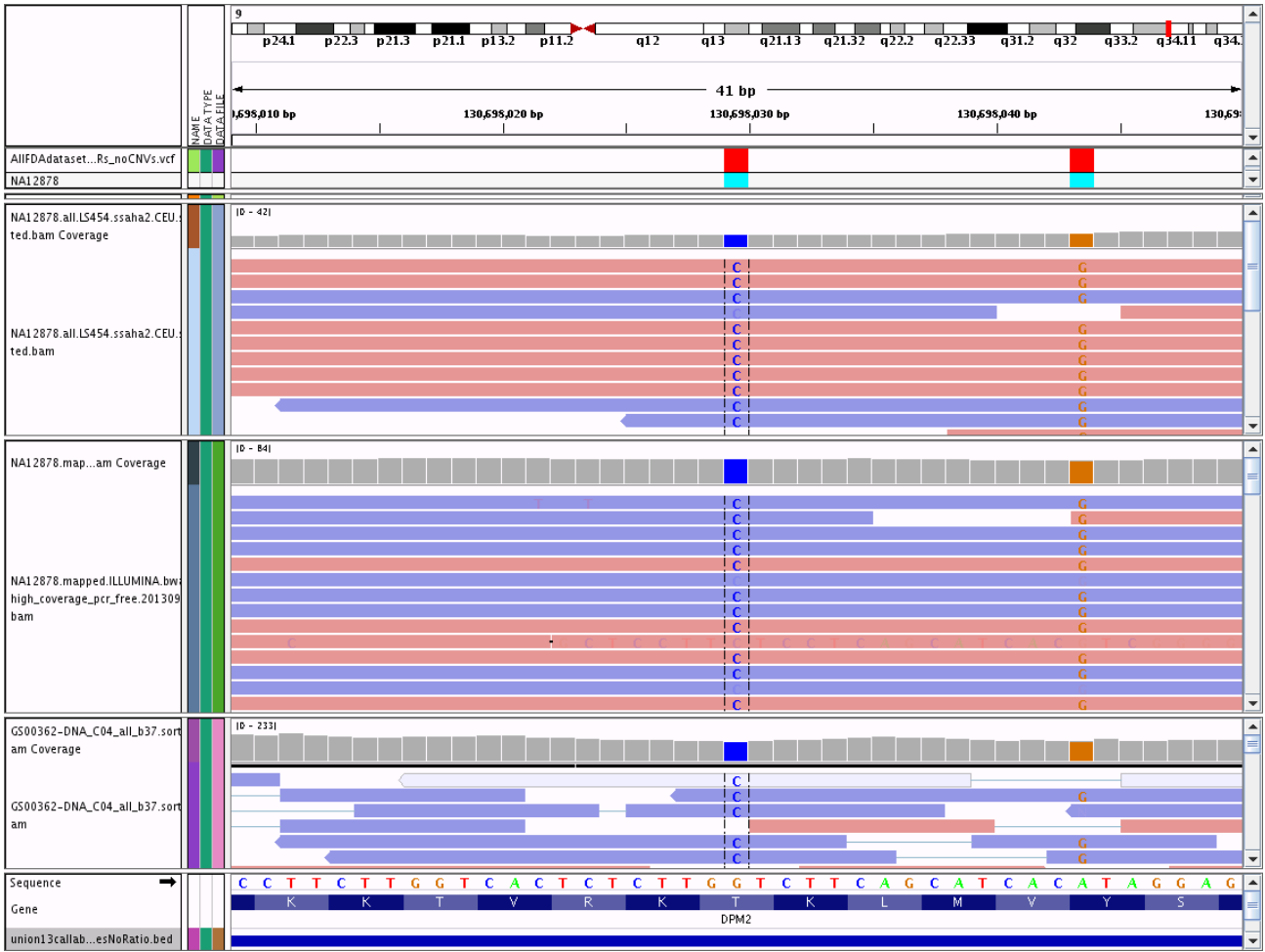

Fig. S27: Example of a location that is called homozygous reference by the OMNI microarray but is actually two nearby SNPs. Displayed from top to bottom are the fosmid vcf, our highly confident vcf, 454 whole genome alignments, 250x250 bp Illumina alignments, Complete Genomics alignments, and our highly confident regions.

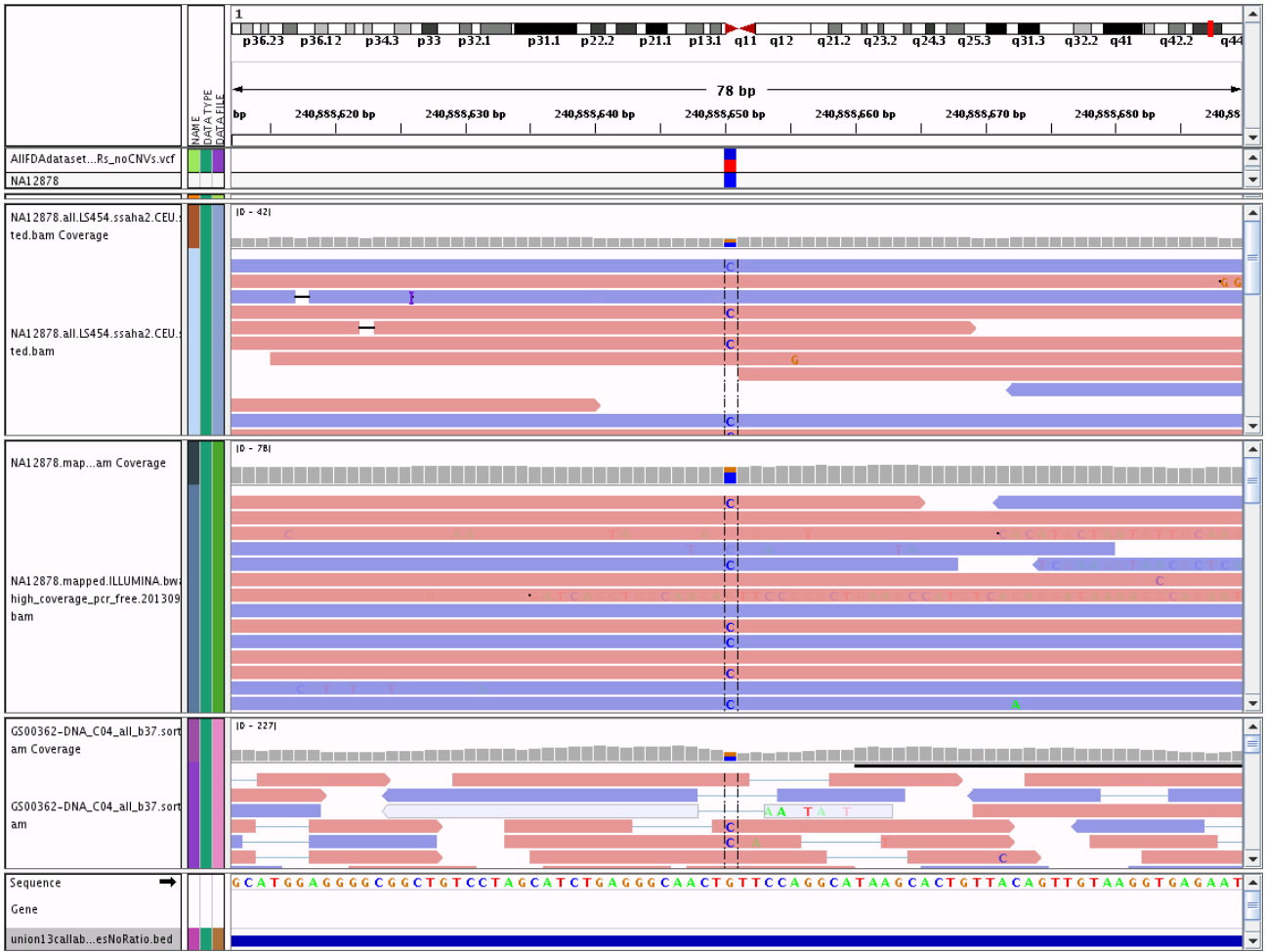

Fig. S28: Example of a location that is called homozygous reference by the OMNI microarray because the probe is for an A SNP and the actual SNP in this sample is a C. Displayed from top to bottom are the fosmid vcf, our highly confident vcf, 454 whole genome alignments, 250x250 bp Illumina alignments, Complete Genomics alignments, and our highly confident regions.

(a) 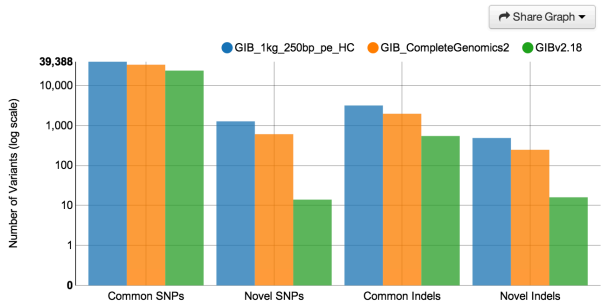

(b) 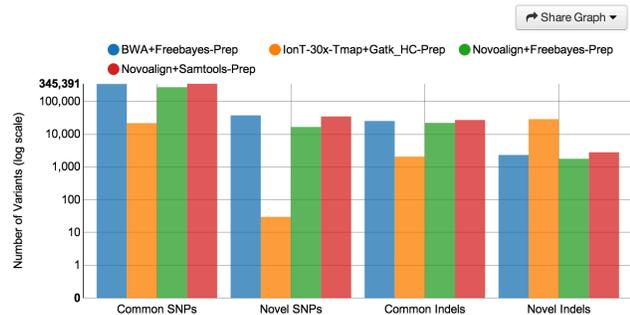

Fig. S29: (a) Summary of ***exome SNPs and indels*** called in our whole genome integrated calls (GiBv2.15b), 250bp whole genome Illumina called with GATK HaplotypeCaller v.2.6 (GiB_1kg_250bp_pe_HC), and whole genome Complete Genomics v2.0 (GiB_CompleteGenomics).  (b) Summary of exome SNPs and Indels called in 150x Illumina exome sequencing mapped with BWA and called with Freebayes (BWA+Freebayes-Prep), 30x Ion Torrent exome sequencing mapped with Tmap and called with GATK HaplotypeCaller (IonT-30x-Tmap+Gatk_HC-Prep), 150x Illumina exome sequencing mapped with Novoalign and called with Freebayes (Novoalign+Freebayes-Prep), and 150x Illumina exome sequencing mapped with Novoalign and called with Samtools (Novoalign+Samtools-Prep).  Note that the variants in the whole genome datasets (a) only include variants in the exome regions, while the variants in the exome datasets (b) include all called variants in this figure.  However, in all other figures, the variants are subsetted to only include variants in the intersection of the exome and highly confident integrated bed files. This figure and Figs. S5-S7 can be generated and modified at (a) http://www.bioplanet.com/gcat/reports/575/variant-calls/genome-in-a-bottle-exome/gib-1kg-250bp-pe-hc/compare-1617-1626/group-read-depth and (b) http://www.bioplanet.com/gcat/reports/577/variant-calls/genome-in-a-bottle-exome/bwa-freebayes-prep/compare-579-558-578/group-read-depth by selecting SNPs and/or indels and the desired "Graph By" for the ROC curves.

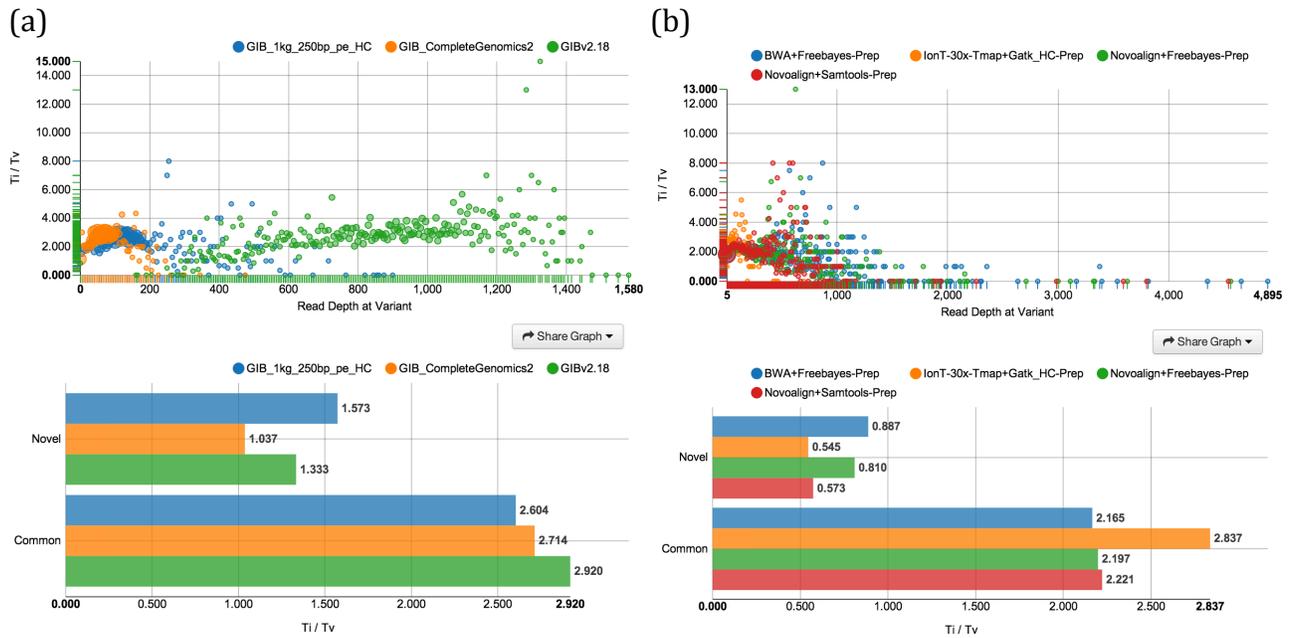

Fig. S30: Transition/transversion ratio (Ti/Tv) for *exome SNPs* for the same datasets in Fig. S5, plotted vs. Read Depth (top) and divided into novel and common variants (bottom).

(a)
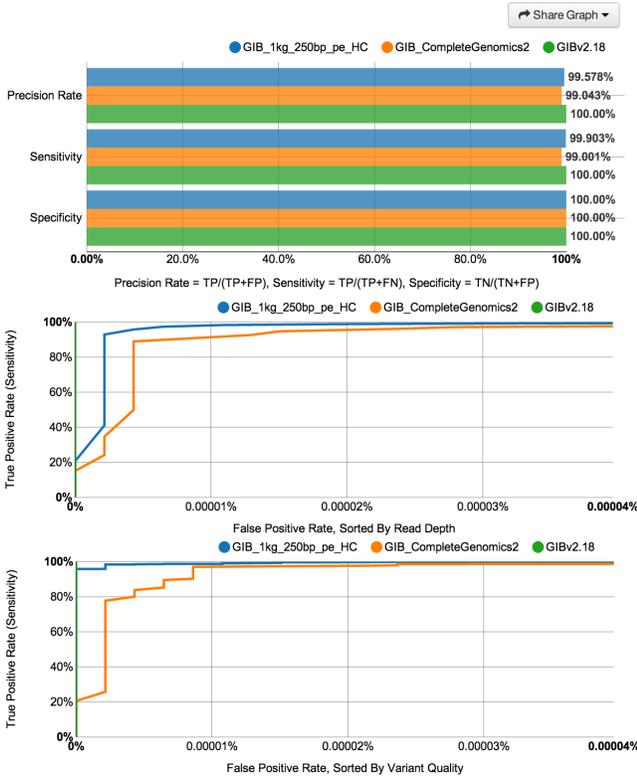

(b)
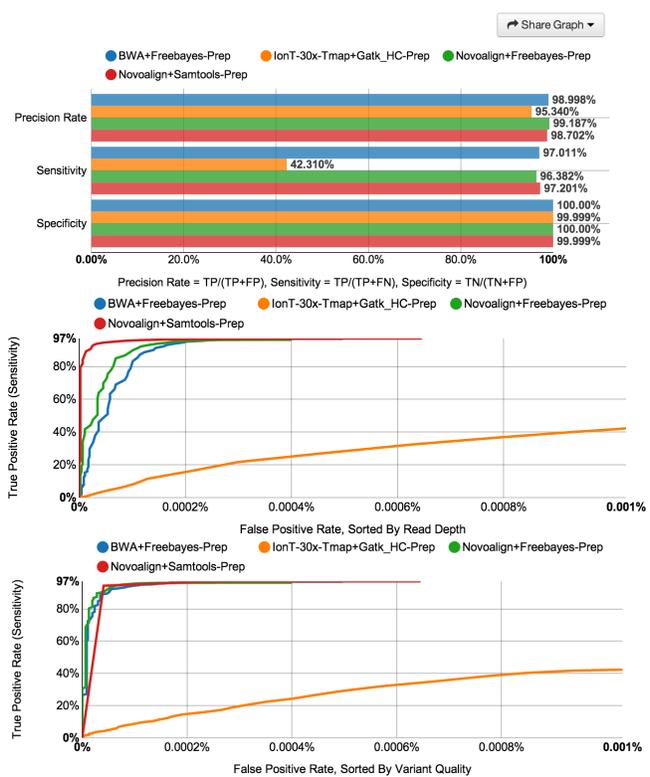

Fig. S31: Performance assessment of *exome SNPs* for the datasets in Fig. S4 using our integrated genotypes as a benchmark, excluding uncertain regions including structural variants in dbVar. The top tables summarize overlap of individual datasets with our integrated genotypes. In the last 4 columns, the genotype of the individual dataset is before the dash, and the genotype of our integrated calls is after the dash. The bar graphs depict the Precision Rate (TP/(TP+FP)), Sensitivity (TP/(TP+FN)), and Specificity (TN/(TN+FP)) for each dataset using our integrated genotypes as a benchmark. Finally, Receiver Operating Characteristic (ROC) curves plotting True Positive Rate vs. False Positive Rate are shown sorted by Read Depth (top) or Variant Quality (bottom).

(a) 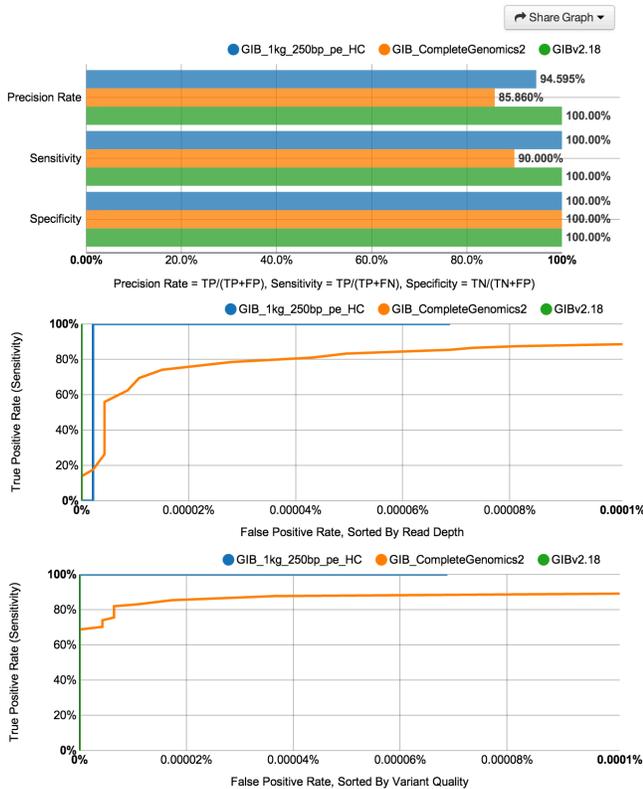

(b) 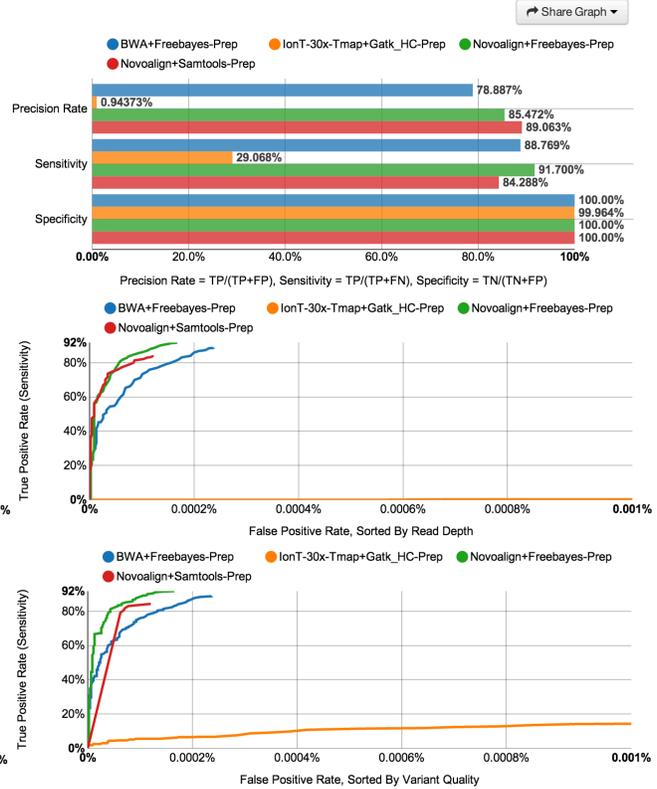

Fig. S32: Performance assessment of ***exome indels*** for the datasets in Fig. S4 using our integrated genotypes as a benchmark, excluding uncertain regions including structural variants in dbVar. The top tables summarize overlap of individual datasets with our integrated genotypes. In the last 4 columns, the genotype of the individual dataset is before the dash, and the genotype of our integrated calls is after the dash. The bar graphs depict the Precision Rate (TP/(TP+FP)), Sensitivity (TP/(TP+FN)), and Specificity (TN/(TN+FP)) for each dataset using our integrated genotypes as a benchmark. Finally, Receiver Operating Characteristic (ROC) curves plotting True Positive Rate vs. False Positive Rate are shown sorted by Read Depth (top) or Variant Quality (bottom).

| Pipeline | GIB Sensitivity | GIB Specificity | Ti/Tv | SNPs | Indels | Novel % |
|---|---|---|---|---|---|---|
| 1kg_250bp_pe_HC | 99.88% | 99.9991% | 2.014 | 3,883,814 (80.28%) | 948,124 (19.60%) | 255,671 (5.28%) |
| CompleteGenomics | 95.75% | 99.9985% | 2.076 | 3,509,397 (88.22%) | 467,271 (11.75%) | 116,484 (2.93%) |
| GIB v2.18 WGS | 100.00% | 100.0000% | 2.136 | 2,728,554 (94.50%) | 158,632 (5.49%) | 11,374 (0.39%) |

> Variant Types

This chart shows the breakdown of the variant classes by SNPs, Insertions, and Deletions.

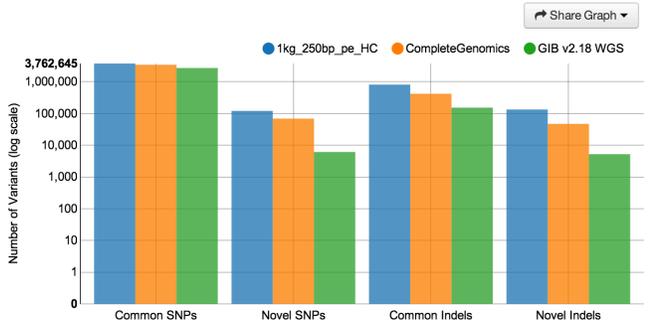

Fig. S33: Summary of *whole genome SNPs and indels* from our whole genome integrated calls (GiB v2.18 WGS), Complete Genomics 2.0 (CompleteGenomics) and 250bp Illumina mapped with BWA-MEM and called with GATK HaplotypeCaller v2.6 (1kg_250bp_pe_HC).

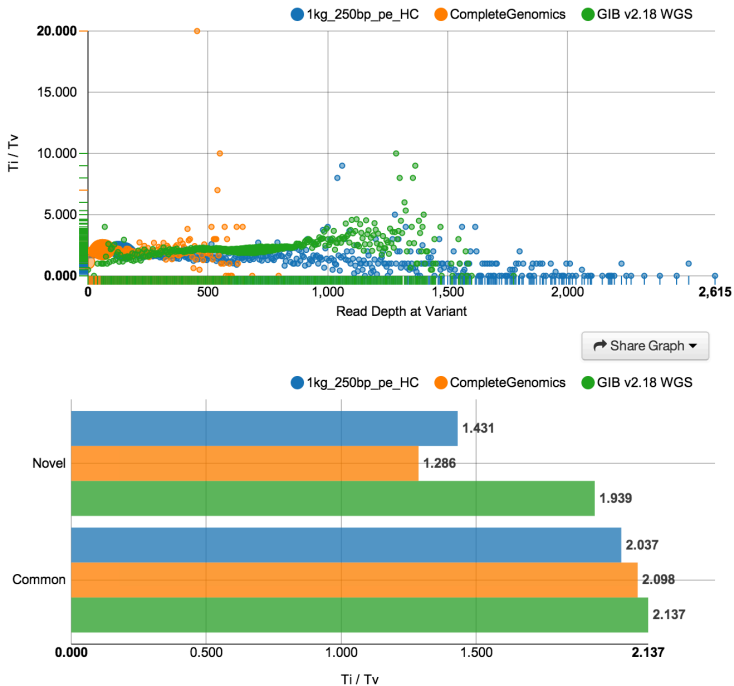

Fig. S34: Transition/transversion ratio (Ti/Tv) for *whole genome SNPs* for the Complete Genomics and 250bp Illumina datasets in Fig. S7, plotted vs. Read Depth (top) and divided into novel and common variants (bottom).

(a)

| Pipeline | True+ | False+ | True- | False- | Het-Ref | Het-HomVar | HomVar-Het | HomVar-Ref |
|---|---|---|---|---|---|---|---|---|
| 1kg_250bp_pe_HC | 2,724,914 | 7,394 | 2,191,888,563 | 2,853 | 6,220 | 695 | 114 | 164 |
| CompleteGenomics | 2,648,506 | 19,528 | 2,191,876,429 | 79,496 | 17,974 | 662 | 56 | 567 |
| GIB v2.18 WGS | 2,728,710 | 0 | 2,191,895,957 | 0 | 0 | 0 | 0 | 0 |

(b)

| Pipeline | True+ | False+ | True- | False- | Het-Ref | Het-HomVar | HomVar-Het | HomVar-Ref |
|---|---|---|---|---|---|---|---|---|
| 1kg_250bp_pe_HC | 154,680 | 11,444 | 2,191,884,513 | 705 | 6,705 | 403 | 87 | 1,277 |
| CompleteGenomics | 113,244 | 13,392 | 2,191,882,565 | 43,460 | 10,822 | 1,316 | 22 | 425 |
| GIB v2.18 WGS | 158,788 | 0 | 2,191,895,957 | 0 | 0 | 0 | 0 | 0 |

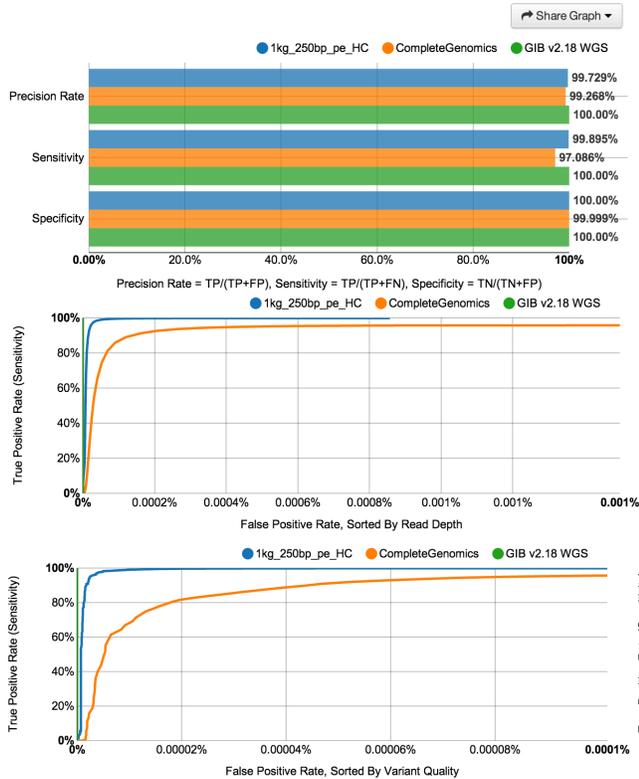
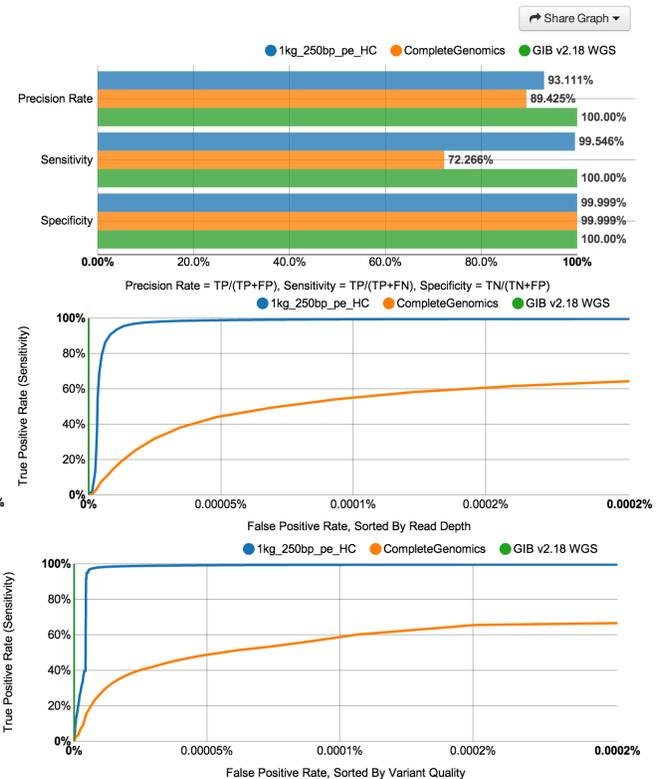

Fig. S35: Performance assessment of **whole genome (a) SNPs and (b) indels** for the Complete Genomics and 250bp Illumina datasets in Fig. S7 using our integrated genotypes as a benchmark, excluding uncertain regions including structural variants in dbVar. The top tables summarize overlap of individual datasets with our integrated genotypes. In the last 4 columns, the genotype of the individual dataset is before the dash, and the genotype of our integrated calls is after the dash. The bar graphs depict the Precision Rate (TP/(TP+FP)), Sensitivity (TP/(TP+FN)), and Specificity (TN/(TN+FP)) for each dataset using our integrated genotypes as a benchmark. Finally, Receiver Operating Characteristic (ROC) curves plotting True Positive Rate vs. False Positive Rate are shown sorted by Read Depth (top) or Variant Quality (bottom).

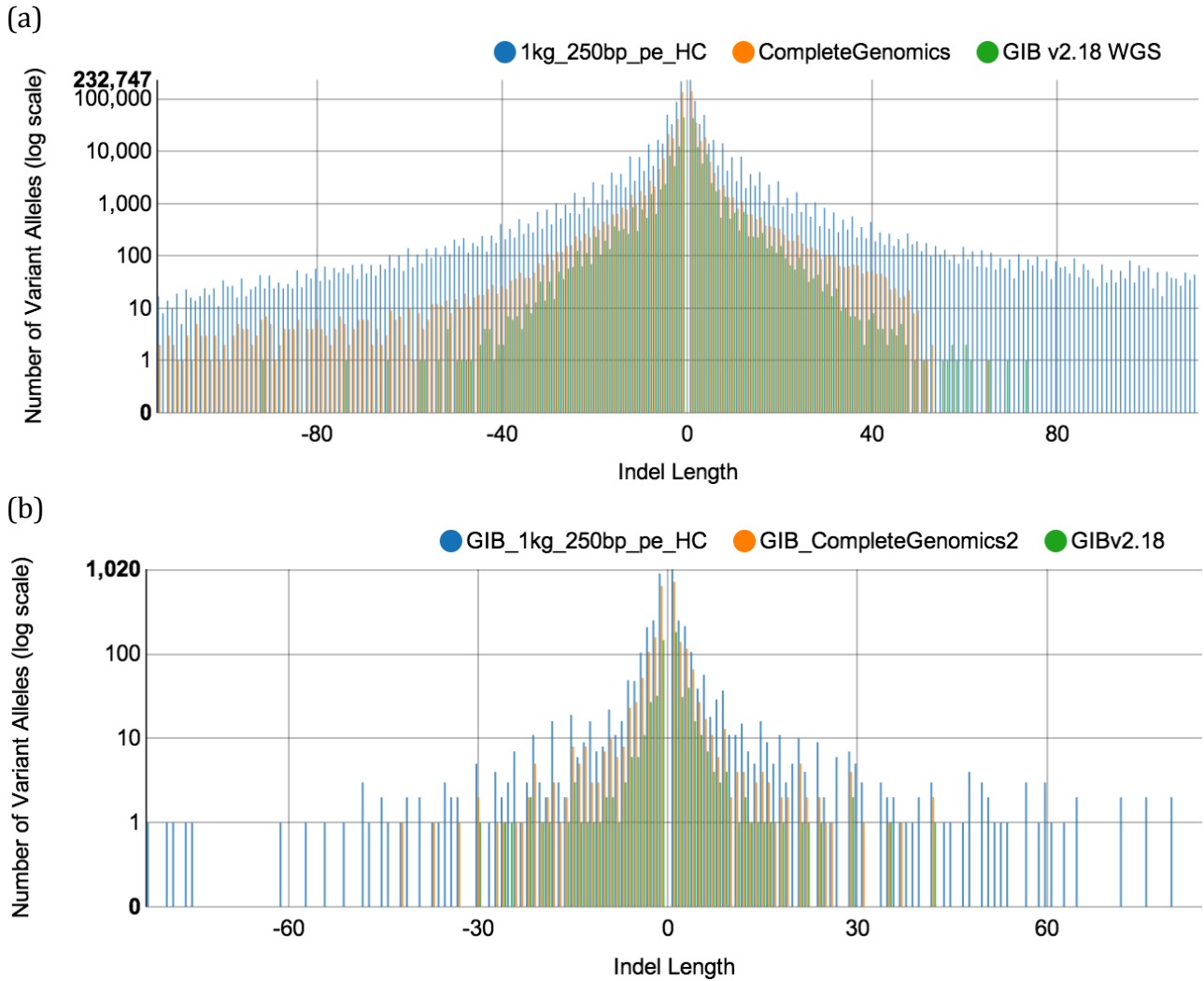

Fig. S36: Indel length distributions in the (a) whole genome and (b) exome. (a) The whole genome indel length distributions are shown for our whole genome integrated calls (green, GiBv2.18), Complete Genomics 2.0 (GIB_CompleteGenomics2) and 250bp Illumina mapped with BWA-MEM and called with GATK HaplotypeCaller v2.6 (GIB_1kg_250bp_pe_HC)

Fig. S37: Example of site (chr1:2843339) that is likely to be homozygous reference, but is called a heterozygous T/C SNP by the 250-bp Illumina sequencing dataset due to an apparent systematic sequencing error that also occurs at a low fraction in the 100-bp Illumina whole genome sequencing (top, CEU) and even in 100-bp Illumina fosmid sequencing, which should only have homozygous variants (bottom, NA12878-7k). Other platforms have (454, Complete Genomics, and SOLiD) have no evidence of a variant at this site.

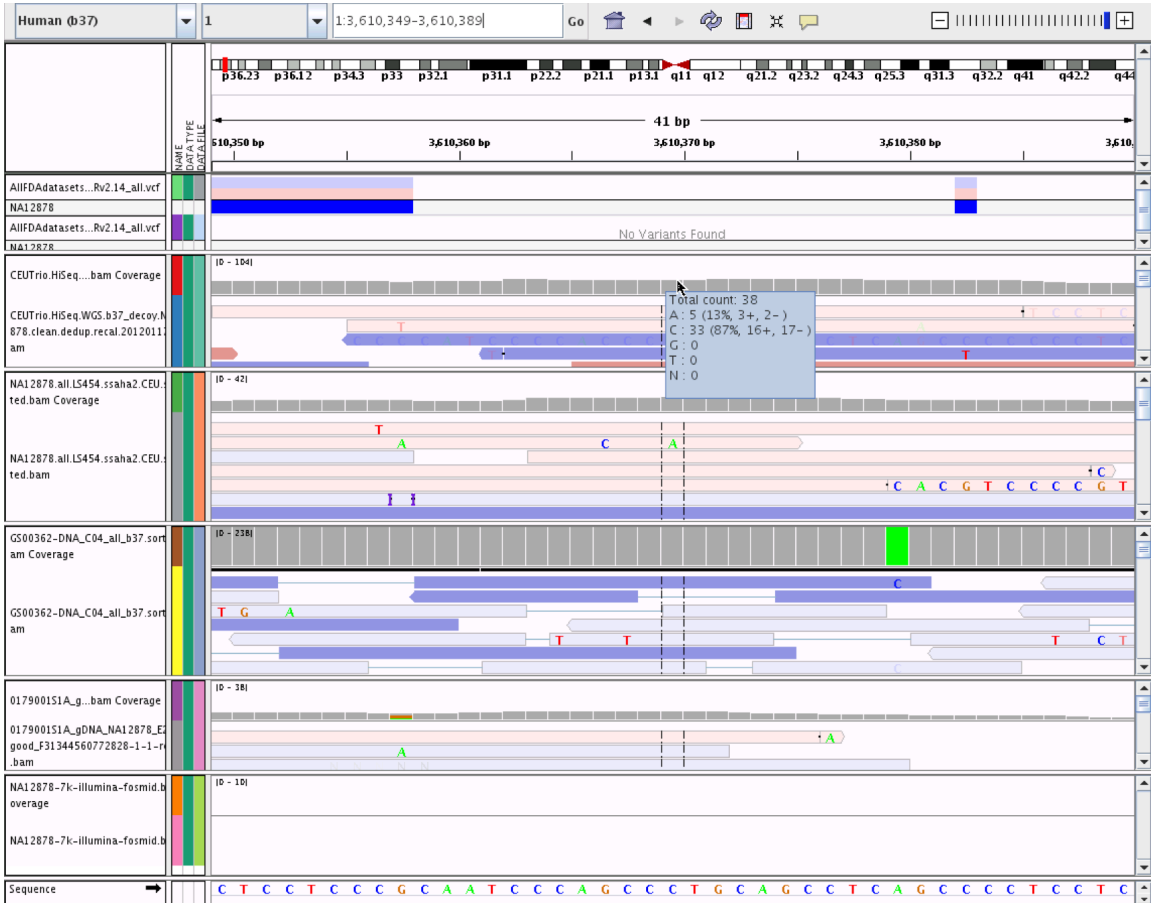

Fig. S38: Example of low mapping quality site where the 250-bp Illumina vcf has a SNP and our integrated genotypes call homozygous reference. Many discordant genotypes fall in this category, where a low fraction of reads contains variants and many reads have low mapping quality, so it is difficult to determine the correct genotype.

41